\documentclass[11pt,a4paper]{article}
\usepackage{jheppub}
\usepackage{amsfonts}
\usepackage{amssymb}
\usepackage{amsmath}

\newcommand{\be}{\begin{equation}}
\newcommand{\ee}{\end{equation}}
\newcommand{\bea}{\begin{eqnarray}}
\newcommand{\eea}{\end{eqnarray}}

\newcommand{\bb}{\bibitem}

\title{Extended Supersymmetry and BPS solutions in baby Skyrme models}
\author[a]{C. Adam,}
\author[a]{J.M. Queiruga,}
\author[a]{J. Sanchez-Guillen}
\author[b]{and A. Wereszczynski}
\affiliation[a]{University of Santiago de Compostela,
Santiago de Compostela, Spain \\ and  Instituto Galego de F\'isica de Altas Enerxias (IGFAE) \\ E-15782 Santiago de Compostela, Spain}
\affiliation[b]{Institute of Physics,  Jagiellonian University,
        Reymonta 4, Krak\'{o}w, Poland}
\emailAdd{adam@fpaxp1.usc.es}
\emailAdd{queiruga@fpaxp1.usc.es}
\emailAdd{joaquin@fpaxp1.usc.es}
\emailAdd{andwereszczynski@gmail.com}
\abstract{
We continue the investigation of supersymmetric extensions of baby Skyrme models in $d=2+1$ dimensions. In a first step, we show that the CP(1) form of the baby Skyrme model allows for the same $N=1$ SUSY extension as its O(3) formulation. Then we construct the $N=1$ SUSY extension of the gauged baby Skyrme model, i.e., the baby Skyrme model coupled to Maxwell electrodynamics. In a next step, we investigate the issue of $N=2$ SUSY extensions of baby Skyrme models. We find that {\em all} gauged and ungauged submodels of the baby Skyrme model which 
support BPS soliton solutions allow for an $N=2$ extension such that the BPS solutions are one-half BPS states (i.e., annihilated by one-half of the SUSY charges). In the course of our investigation, we also derive the general BPS equations for completely general $N=2$ supersymmetric field theories of (both gauged and ungauged) chiral superfields, and apply them to the gauged nonlinear sigma model as a further, concrete example.}
\arxivnumber{1234.5678}

\begin{document}
\maketitle

\section{Introduction}

The Skyrme model \cite{skyrme} is a nonlinear field theory in 3+1 dimensional Minkowski space which supports topological soliton solutions.  Its field variables take values in SU(2) which, together with an one-point compactification of the base space ${\mathbb R}^3 \to S^3$ implied by the condition of finite energy, leads to the classification of field configurations by an integer-valued winding number or topological degree. The most important application of the Skyrme model is in the field of nuclear and strong interaction physics \cite{AdNaWi}-\cite{man-sut-book}. In this context, the Skyrme model (or some of its generalizations) is interpreted as a low energy effective field theory which may be justified from the underlying fundamental theory (QCD), e.g., by invoking some large $N_c$ (number of colors) arguments \cite{thooft}, \cite{witten1}. In this interpretation, the primary fields of the effective theory are related to mesons (e.g. pions in the SU(2) case), whereas baryons and nuclei are described by the topological solitons of the theory, and baryon number is identified with the topological degree of the corresponding soliton.  

The baby Skyrme model was introduced originally as a planar analogue of the three dimensional
Skyrme model \cite{old}-\cite{foster1}, although it has found its own applications, e.g., in condensed matter physics \cite{qhe} or 
in brane cosmology \cite{sawa1}. Its target space is simplified accordingly, as well ($S^2$
instead of the $SU(2)$ target space of the Skyrme model), such that static field configurations
again can be classified by a winding number. 
Like the original version of the
Skyrme model, as proposed by Skyrme, also the Lagrangian of the baby Skyrme model consists of a kinetic term quadratic in first derivatives (the
$O(3)$ nonlinear sigma model term) and a quartic kinetic term (the analogue of the Skyrme
term). Further, for the baby Skyrme model, the inclusion of a potential term is
obligatory for the existence of static finite energy solutions. The specific form of this
potential term is, however, quite arbitrary, and different potentials have been studied
\cite{old}-\cite{foster1}. The Skyrme model, too, allows for the addition of a potential (not obligatory in that case) or of some further terms like, e.g., the square of the topological current, which is sextic in first derivatives.  
In any case, the presence of higher derivative terms ("non-standard kinetic terms")  in Skyrme-type models is necessary for the existence of topological solitons.
In addition, in both models the energies of static configurations can be bound from below by a Bogomol'nyi bound (a multiple of the topological degree), but generic soliton solutions do not saturate this bound. It is, however, possible both for the baby Skyrme model 
\cite{GP}-\cite{Sp1}
and for a generalized Skyrme model 
\cite{BPS-Sk1}-\cite{BoMa}
(i.e., a generalization of the original model proposed by Skyrme) to find certain submodels such that their topological soliton solutions saturate the corresponding Bogomol'nyi bound, that is, they are of the BPS type and obey  certain first-order BPS equations.   

At this point, it is useful to compare the properties of Skyrme-type theories with those of the abelian and nonabelian Higgs models with their vortex-type or monopole-type solitons (see e.g. \cite{man-sut-book}). The topology of these solitons is different, because now the one-point compactification of the base space is not assumed, and the scalar fields ("Higgs fields") may be classified by a topological degree related to their winding about the sphere at spatial infinity. As mentioned already, this behaviour leads to an infinite energy due to the presence of angular gradients in the kinetic energy density with a rather slow decay for large distances. The well-known way to remedy this problem is via the coupling of the Higgs field to a gauge field such that the unwanted angular gradients are converted into pure gauge configurations and do not contribute to the energy. If the standard kinetic terms for the gauge fields (Maxwell or Yang-Mills terms) are added, we just arrive at the abelian Higgs model or the t'Hooft-Polyakov (nonabelian) Higgs model, respectively. In spite of the different topology of the corresponding solitons, these theories share many properties with the Skyrme-type ones. For the Higgs theories, too, the energies of static configurations can be bound from below by Bogomol'nyi bounds where, however, solitons of generic theories do not saturate the bounds. Again, submodels can be found (usually, by a judicious choice of the Higgs potential) whose solitons (vortices or monopoles) are of the BPS type and saturate the bound.  

There exists, however, one aspect where the two classes of theories are apparently rather different, namely the issue of supersymmetry. The Higgs-type theories are well-known to possess $N=1$ supersymmetric (SUSY) extensions. Further, the submodels with BPS solitons even allow for an $N=2$ SUSY extension such that the BPS soliton solutions are, in fact, one-half BPS states in the sense of SUSY, that is, field configurations which are annihilated by one-half of the SUSY charges (see, e.g., \cite{divecchia-ferrara} - \cite{edelstein-nunez}). The construction of these SUSY extensions is facilitated by the fact that the kinetic terms both for the Higgs and for the gauge fields are of the standard form (quadratic in derivatives), because the SUSY extensions of these kinetic terms are well-known. 
On the other hand, until recently not much was known about the SUSY extensions of Skyrme-type theories, where the presence of nonstandard kinetic terms is mandatory. To the best of our knowledge, the first investigations of SUSY extensions of Skyrme-type theories were performed in
\cite{nepo}, \cite{frey}. Concretely, the authors studied possible SUSY extensions of the so-called Skyrme-Faddeev-Niemi (SFN) model, which has exactly the field content of the baby Skyrme model, but in 3+1 dimensions (in 3 spatial dimensions the potential term is not mandatory and is usually omitted). In this model field configurations are no longer classified by a winding number but, instead, by a linking number (the Hopf index). In both papers, the authors treated the SFN model  as a CP(1) restriction of the original Skyrme model, where the elimination of the third, unwanted degree of freedom is achieved by transforming it into pure gauge via the introduction of a non-dynamical gauge field. One consequence of this procedure is that the Skyrme term (which is non-standard) may be expressed as the standard Maxwell term of the non-dynamical gauge field. As a consequence, the resulting action only contains standard kinetic terms (the nonlinear sigma model term for the CP(1) field and the Maxwell term) and standard SUSY techniques may be used. The result of these investigations is that the original SFN model cannot be extended to a SUSY theory by these methods. Any SUSY extension achieved in this way contains additional terms already in the bosonic sector. 

The investigation of SUSY extensions of genuinely nonstandard kinetic terms has been resumed only recently \cite{bazeia-susy}-\cite{FarKe2} (see also \cite{nitta1}-\cite{adel1} for related discussions), where this rising interest is partly owed to the fact that field theories with non-standard kinetic terms may be instrumental in the resolution of some enigmas of cosmology \cite{k-infl}-\cite{Dzhu1}. 
Concretely, in \cite{susy-bS} we demonstrated that the baby Skyrme model in the O(3) formulation does have a $N=1$ SUSY extension for arbitrary non-negative potential. It turned out, however, that a submodel supporting BPS solitons (the so-called BPS baby Skyrme model, where the non-linear sigma model term is suppressed) cannot be supersymmetrically extended by the methods of that paper. It is the purpose of the present paper to go much further in the analysis of SUSY extensions of baby Skyrme models where, among other issues, the puzzle just mentioned will be resolved in the course of the investigation.  

Our paper is organized as follows. In Section 2 we introduce the baby Skyrme models and fix some notation. In Section 3 we give our conventions for $N=1$ SUSY in 2+1 dimensions. In Section 4 we discuss the $N=1$ SUSY extension of the baby Skyrme model in the CP(1) formulation. In Section 5 we introduce the $N=1$ SUSY extension of the gauged baby Skyrme model \cite{GPS}, i.e., the baby Skyrme model coupled to Maxwell electrodynamics in the standard way. In Section 6 we give our conventions for extended $N=2$ SUSY. In Section 7 we attempt to find an $N=2$ extension of the SUSY baby Skyrme model. We find that, in addition to the well-known 
Kaehler potential term giving rise to the non-linear sigma model, we have to introduce a further term into the lagrangian superfield. This further term has the surprising effect that, after the substitution of the auxiliary fields via their field equations, not only the quartic (Skyrme) term is produced in the bosonic sector but, at the same time, the quadratic (nonlinear sigma model) term is eliminated for arbitrary values of the Kaehler potential. Besides, a potential term depending on the Kaehler metric is automatically induced in this process. In other words, in the purely bosonic sector we find precisely the BPS baby Skyrme model consisting of the Skyrme term and a potential, but without the sigma model term. Due to the absence of this sigma model term, we may choose arbitrary Kaehler metrics and, therefore, arbitrary non-negative potentials. So in this case the potential is induced by the Kaehler metric and {\em not} by a superpotential. A superpotential term is, in fact, forbidden in this construction. 
In Section 8 we discuss the issue of BPS (or Bogomol'nyi) equations for the BPS baby Skyrme models from the point of view of $N=2$ SUSY. 
Concretely, in a first step we derive the equation for one-half BPS states for a {\em completely general} $N=2$ chiral superfield. Then we apply the resulting equation to the SUSY BPS baby Skyrme model and find that its one-half BPS states are precisely the BPS solutions of the BPS baby Skyrme model \cite{GP}-\cite{Sp1}. In Section 9 we introduce the $N=2$ SUSY extension of the gauged baby Skyrme model. Again, the procedure implies the absence of the (gauged) quadratic sigma model term, and we find the gauged BPS baby Skyrme model \cite{gaugedBPS-bS}. For this model, a BPS bound and BPS solitons have been found recently, where the construction of the BPS bound implied the introduction of a certain "superpotential" ${\cal W}$ which is related to the potential ${\cal V}$ by a first order differential equation ("superpotential equation"). We find that, again, the BPS solitons are one-half BPS states of the $N=2$ SUSY extension, and the "superpotential equation" may be understood from the fact that both the "superpotential" ${\cal W}$ and the potential ${\cal V}$ are derived from a certain Kaehler potential. In Section 10 we apply our methods to the gauged nonlinear sigma model, which is known to possess BPS solitons for a certain choice of potential \cite{schr1}. It follows easily from our general construction that this model has an $N=2$ SUSY extension and that the BPS solitons are one-half BPS states. In this case, the sigma model term and, therefore, the Kaehler metric, have a fixed, given form, so, as a result, also the potential (which is again a function of the Kaehler metric) is fixed. Finally, Section 11 contains our conclusions. 
  
\section{The baby Skyrme model}
The field variables of the baby Skyrme model take values in the two-sphere, so it is naturally parametrized by a three component unit vector field $\vec n (x)$, where $\vec n^2 =1$. The lagrangian density is a sum of three terms,
\be
L^{bS} = L_2 + L_4 + L_0
\ee
where ${ L}_2$ is the sigma model term
\be
{ L}_2 = \frac{\lambda_2}{4}(\partial_\mu \vec n )^2 ,
\ee
${ L}_4$ is the Skyrme term
\be
{ L}_4 = -\frac{\lambda_4}{8}(\partial_\mu \vec n \times \partial_\nu \vec n )^2 \equiv -\frac{\lambda_4}{16} K_\mu^2  ,
\ee
where $K^\mu$ is the topological current
\be
K^\mu = \epsilon^{\mu\nu\rho} \vec n \cdot (\partial_\nu \vec n \times \partial_\rho \vec n ) ,
\ee
such that 
\be
k=(1/8\pi) \int d^2 x K^0, \quad k \in {\mathbb Z}
\ee
is the winding number (topological degree) of the map $\vec n$. Finally, 
 ${ L}_0$ is the potential term
\be
{ L}_0 = -\lambda_0 {\cal V}(\vec n  ).
\ee
The lagrangian density has dimensions of $\frac{[\mbox{action}]}{[\mbox{length}]^2 [\mbox{time}]}$ or, equivalently, 
$\frac{[\mbox{energy}]}{[\mbox{length}]^2}$. Further, we shall assume natural units where the velocity of light is equal to one such that 
$[\mbox{length}] = [\mbox{time}]$. 
Extracting a common energy scale $E_0$ we may write the Lagrangian density like 
\be \label{bS-lag}
L=E_0  \left( \frac{\nu^2}{4} (\partial_\mu \vec n )^2 - \frac{\lambda^2}{8} (\partial_\mu \vec n \times \partial_\nu \vec n )^2 
-\mu^2 {\cal V}(\vec n )  \right) 
\ee
where now $\nu$ is dimensionless, and $\lambda$ and $\mu^{-1}$ have the dimension of length. A nonzero $\nu$ may always be set equal to one, $\nu =1$, by an appropiate choice of the energy scale $E_0$. We shall, therefore, assume $\nu =1$ or $\nu =0$ in what follows, depending on whether the term $L_2$ is present or absent. Besides, all energies will be measured in units of $E_0$, which is equivalent to setting $E_0 =1$, what we assume from now on. In a next step, we shall introduce dimensionless coordinates via $x^\mu = l_0 y^\mu$ (here, $l_0$ is a universal length scale) which are more appropriate for SUSY calculations, where we continue, however, to use the symbols $x^\mu$ (instead of $y^\mu$) for the new, dimensionless coordinates. For nonzero $\lambda$, we may always choose $l_0 = \lambda$. Choosing, in addition, length units such that $l_0=1$, we get again the lagrangian (\ref{bS-lag}) where, now, both $\nu$ and $\lambda$ take the values 1 or 0 (depending on whether the corresponding terms are present or absent), and $\mu$ is a dimensionless coupling constant. But we have not yet made any assumption on the form of ${\cal V}$, therefore we may always reabsorb this constant into the definition of the potential. Doing so, our lagrangian density for the full baby Skyrme model (with all terms present) now reads 
\be \label{bS-lag-2}
L=  \left( \frac{1}{4} (\partial_\mu \vec n )^2 - \frac{1}{8} (\partial_\mu \vec n \times \partial_\nu \vec n )^2 
- {\cal V}(\vec n )  \right) ,
\ee
which is the dimenionless lagrangian density in the O(3) formulation of the baby Skyrme model. In the following, however, we shall need the model in the CP(1) formulation, where the field variable is parametrized by a complex scalar field $u(x)$ related to $\vec n$ by stereographic projection,
\begin{equation}
\vec{n}=\frac{1}{1+\vert u \vert^2}(u+\bar{u},-i(u-\bar{u}), 1 - \vert u \vert^2 ) .
\end{equation}
In terms of the field $u$ (i.e., in CP(1) formulation), the dimensionless lagrangian density reads 
\begin{equation} \label{bS-cp1-lag}
L_{{C}P^1}^{bS}=L_2+L_4+L_0
\end{equation}
where 
\begin{equation}
L_2=\frac{\partial_\mu u \partial^\mu \bar{u}}{(1+u\bar{u})^2}
\end{equation}
is the nonlinear sigma-model term,
\begin{equation}
L_4=-\frac{1}{(1+u\bar{u})^4}\lbrack (\partial_\mu u \partial^\mu \bar{u})^2- (\partial_\mu u \partial^\mu u)(\partial_\nu \bar{u} \partial^\nu \bar{u})   \rbrack
\end{equation}
is the "Skyrme" term quartic in first derivatives, and 
\begin{equation}
L_0=-{\cal V}(u\bar{u})
\end{equation}
is the potential term. From now on, we assume that ${\cal V}$ only depends on the modulus (squared) of $u$ (i.e., only depends on $n_3$ in the O(3) formulation), which implies that the potential does not completely break the SU(2) target space symmetry (the O(3) symmetry in the O(3) formulation) of $L_2 + L_4$, but leaves a U(1) subgroup  (the phase transformation $u\to e^{i\lambda }u$) intact. This is of special importance if we want to couple the Skyrme field $u$ to the U(1) gauge field of electrodynamics. 

It is sometimes useful to consider the slightly more general class of models given by   
\begin{equation}
L_2=g(u ,\bar u)\partial_\mu u \partial^\mu \bar{u},
\end{equation}
\begin{equation}
L_4= -h(u ,\bar u) \lbrack (\partial_\mu u \partial^\mu \bar{u})^2- (\partial_\mu u \partial^\mu u)(\partial_\nu \bar{u} \partial^\nu \bar{u})   \rbrack
\end{equation}
where the original baby Skyrme model corresponds to the choice
\begin{equation}
g(u,\bar{u})^2 = h(u , \bar u) =\frac{1}{(1+u\bar{u})^4}.
\end{equation}
Geometrically, $g$ and $h$ may be interpreted as the target space metric and the (square of the) target space area density, respectively.

\section{$N=1$ supersymmetry in $d=2+1$ dimensions}

We use the Minkowski space metric $\eta_{\mu\nu} = {\rm diag} (+,-,-)$. Then, an $N=1$ real scalar superfield is given by
\begin{equation}
\Phi (z)=\phi (x)+\theta^\alpha\psi_\alpha (x)-\theta^2F(x), 
\end{equation}
where the coordinate $z$ stands collectively for $(x^\mu,\theta_\alpha)$, $\phi$ is a real scalar field, $\psi_\alpha $ is a fermionic two-component Majorana spinor, and $F$ is the auxiliary field. Further, $\theta^\alpha$ are the two Grassmann-valued superspace coordinates,
and $\theta^2 \equiv (1/2)\theta^\alpha \theta_\alpha$. 
The components of a superfield can be extracted with the help of the following projections
\be
\label{comp}
\phi(x)=\Phi(z)|,\quad\,\psi_{\alpha}(x)=D_{\alpha}\Phi(z)|,\quad\,F(x)=D^2\Phi(z)|,
\ee
where the superderivative is
\be
D_\alpha = \frac{\partial}{\partial\theta^\alpha} -i \gamma^\mu{}_\alpha{}^\beta
\theta_\beta \partial_\mu 
\ee
\be
D^2 \equiv \frac{1}{2} D^\alpha D_\alpha 
\ee
and the vertical line $|$ denotes evaluation at $\theta^\alpha =0$. From here it is easy to construct supersymmetric lagrangian, which are just the $\theta  $ integrals of general superfields, that is, general functions of the basic superfields and their superderivatives, i.e.:
\be
L_{N=1}=\int d^2 \theta {\cal L}(\Phi^i,D^\alpha \Phi^j,...) .
\ee
It is also possible to construct $N=1$ complex superfields by combining real ones.

\section{$N=1$ CP(1) baby Skyrme model}

\subsection{ $N=1$ extension}

 We will construct a $N=1$ supersymmetric extension of the model (\ref{bS-cp1-lag}). In a first step,  we need the basic $N=1$ superfields
\begin{equation}
\Phi^1=\phi^1+\theta^\alpha \psi^1_\alpha-\theta^2 F^1,\quad \Phi^2=\phi^2+\theta^\alpha \psi^2_\alpha-\theta^2 F^2 .
\end{equation}
Taking into account that $u\in \mathbb{C}$ and $\phi^i\in \mathbb{R}$, we introduce the following combinations for the new superfields $U$ and $\bar{U}$:
\begin{eqnarray}
U=\Phi^1+i\Phi^2\\
\bar{U}=\Phi^1-i\Phi^2
\end{eqnarray}
such that
\begin{eqnarray}
U\vert=\phi^1+i\phi^2\equiv u\\
\bar{U}\vert=\phi^1-i\phi^2\equiv\bar{u} .
\end{eqnarray}
Similarly we define
\begin{eqnarray}
\chi_\alpha\equiv\psi^1_\alpha+i \psi^2_\alpha , &\quad& F\equiv F^1+iF^2\\
 \bar{\chi}_\alpha\equiv\psi^1_\alpha-i \psi^2_\alpha , &\quad& \bar{F}\equiv F^1-iF^2 .
\end{eqnarray}
With these complex combinations of real superfields we can now generate the quadratic term. Considering only the bosonic sector, we find
\begin{equation}
{L}_2\vert_{bos}=\frac{1}{2}\int d^2\theta g(U,\bar{U})D^\alpha U  D_\alpha\bar{U}\vert_{\chi =0} =g(u,\bar{u})(F\bar{F}+\partial_\mu u \partial^\mu \bar{u})
\label{l2}
\end{equation}
that is, the quadratic term of the baby Skyrme model plus a term quadratic in the auxiliary field $F$. For the quartic term we need two contributions, which for the moment we write without their target space area factors $h$. The first one is
\begin{eqnarray}
\tilde{L}_{4a}&=&\int d^2\theta \lbrack  D^\alpha U D_\alpha U + D^\alpha \bar{U} D_\alpha \bar{U}   \rbrack   \lbrack  D^2 U D^2 U +D^2 \bar{U} D^2 \bar{U} -\\
&-&  \frac{1}{4}(D^\alpha D^\beta UD_\alpha D_\beta U+D^\alpha D^\beta \bar{U} D_\alpha D_\beta \bar{U} ) \rbrack 
\end{eqnarray}
and its bosonic part results in
\begin{equation}
\tilde{L}_{4a}\vert_{bos}=(F^2+\bar{F}^2)^2-(\partial_\mu u)^2(\partial_\nu u )^2-(\partial_\mu \bar{u})^2(\partial_\nu \bar{u} )^2-2(\partial_\mu u)^2(\partial_\nu \bar{u} )^2
\end{equation}
For the second contribution 
 we define $A_1=U$ and $A_2=\bar{U}$, then the other part for the quartic Lagrangian is
\begin{equation}
\tilde{L}_{4b}=\sum_{ij}\int d^2\theta \lbrack  D^\alpha A^i D_\alpha A^j   \rbrack   \lbrack  D^2 A^i D^2 A^j -  \frac{1}{4}(D^\alpha D^\beta A^iD_\alpha D_\beta A^j) \rbrack 
\end{equation}
and the bosonic part results in
\begin{equation}
\tilde{L}_{4b}\vert_{bos}= (F^2+\bar{F}^2)^2-(\partial_\mu u)^2(\partial_\nu u )^2-(\partial_\mu \bar{u})^2(\partial_\nu \bar{u} )^2-2(\partial_\mu u \partial^\mu \bar{u})^2
\end{equation}
finally
\begin{equation}
\tilde{L}_{4}\vert_{bos}=-\frac{1}{2}(\tilde{L}_{4a}\vert_{bos}-\tilde{L}_{4b}\vert_{bos})=(\partial_\mu u \partial^\mu \bar{u})^2-(\partial_\mu u)^2(\partial_\nu \bar{u} )^2 .
\end{equation}
We remark for later use that in this specific linear combination, together with the unwanted terms depending on $\partial_\mu u$, also the auxiliary fields $F$ and $\bar F$ have disappeared.

Including now the $h(u,\bar{u})$  factor, we get
\begin{eqnarray}
L_{4a}&=&\int d^2\theta h(U,\bar{U}) \lbrack  D^\alpha U D_\alpha U + D^\alpha \bar{U} D_\alpha \bar{U}   \rbrack   \lbrack  D^2 U D^2 U +D^2 \bar{U} D^2 \bar{U} -\\
&-&  \frac{1}{4}(D^\alpha D^\beta UD_\alpha D_\beta U+D^\alpha D^\beta \bar{U} D_\alpha D_\beta \bar{U} ) \rbrack ,
\end{eqnarray}
\begin{equation}
L_{4b}=\sum_{ij}\int d^2\theta h(U,\bar{U}) \lbrack  D^\alpha A^i D_\alpha A^j   \rbrack   \lbrack  D^2 A^i D^2 A^j -  \frac{1}{4}(D^\alpha D^\beta A^iD_\alpha D_\beta A^j) \rbrack ,
\end{equation}
and the final result for the quartic term has the following form,
\begin{equation} \label{N1-L4}
L_4\vert_{bos}= -h(u,\bar{u}) \lbrack  (\partial_\mu u \partial^\mu \bar{u})^2-(\partial_\mu u)^2(\partial_\nu \bar{u} )^2  \rbrack
\end{equation}
or
\begin{equation}
L_4\vert_{bos}= -\frac{1}{(1+u\bar{u})^4} \lbrack  (\partial_\mu u \partial^\mu \bar{u})^2-(\partial_\mu u)^2(\partial_\nu \bar{u} )^2  \rbrack .
\end{equation}

As usual, in $N=1$ SUSY a potential term results from a (real) superfield ${\cal U}(U ,\bar U)$ called superpotential, which only depends on the basic superfields $U$ and $\bar U$,
\begin{equation}
L_{\cal U}= \int d^2 \theta {\cal U}(U,\bar{U})
\end{equation}
with the bosonic part
\begin{equation}
L_{{\cal U},{bos}}={\cal U}_u F+{\cal U}_{\bar{u}}\bar{F} .
\label{lp}
\end{equation}
Taking into account (\ref{l2}) and (\ref{lp}), the equations of motion for the auxiliary fields are
\begin{equation}
g(u,\bar{u})\bar{F}+{\cal U}_u=0,\quad g(u,\bar{u})F+{\cal U}_{\bar{u}}=0
\end{equation}
or
\begin{equation}
\bar{F}=-\frac{{\cal U}_u}{g(u,\bar{u})},\quad F=-\frac{{\cal U}_{\bar{u}}}{g(u,\bar{u})} .
\end{equation}
Inserting these values in the total Lagrangian we obtain
\begin{eqnarray}
L_{tot}&=&\frac{1}{(1+u\bar{u})^2} \partial_\mu u \partial^\mu \bar{u} - \frac{1}{(1+u\bar{u})^4}\lbrack (\partial_\mu u \partial^\mu \bar{u})^2-\\
&-&(\partial_\mu u \partial^\mu u)(\partial_\nu \bar{u} \partial^\nu \bar{u})  \rbrack - (1+u \bar{u})^2 {\cal U}_u {\cal U}_{\bar{u}}
\end{eqnarray}
and, therefore, precisely the lagrangian density (\ref{bS-cp1-lag}) of the baby Skyrme model with the potential
\be
{\cal V}(u ,\bar u) = (1+u\bar u)^2 {\cal U}_u {\cal U}_{\bar u}.
\ee
For potentials ${\cal V}(u\bar u)$ with the residual U(1) symmetry we have to assume that also ${\cal U}={\cal U}(U\bar U)$.  

What is interesting here is that we cannot eliminate the quadratic term.  Setting $g(u,\bar{u})=0$ at the end leads to
\be
 \frac{{\cal U}_u {\cal U}_{\bar{u}}}{g(u,\bar{u})}\rightarrow \infty ,
\ee
and starting without the quadratic term from the beginning has the consequence that the auxiliary fields only appear linearly in the lagrangian from the superpotential, $L(F,\bar F) \sim {\cal U}_u F+{\cal U}_{\bar{u}}\bar{F} $. They act, therefore, like Lagrange multipliers enforcing the "constraints" 
${\cal U}_u = {\cal U}_{\bar u} =0$. We conclude that, although the quartic term $L_4$ alone can be supersymmetrically extended by the methods of this section, this is not true for the BPS baby Skyrme model $L_4 + L_0$. We shall find in the next section, however, that we may find more general $N=1$ extensions which are capable of producing the BPS Skyrme model in its bosonic sector. Later on, we will see that (reflecting its BPS nature) the BPS baby Skyrme  model even allows for an $N=2$ SUSY extension. In both cases, the potential term $L_0$ is {\em not} induced by a superpotential but, instead, by the target space metric $g$ or by a Kaehler potential related to $g$. 

To summarize the results of this section, for the CP(1) version of the baby Skyrme model we found exactly the same $N=1$ SUSY extension as for its O(3) version \cite{susy-bS}. The two versions are, of course, classically equivalent. The SUSY extensions, however, require the introducion of fermions which must be treated as quantum objects to provide the correct SUSY algebra. The equivalence of the two SUSY extensions is, therefore, not completely obvious, but turns out to be true.

\subsection{More general $N=1$ extensions}

The $N=1$ SUSY extension of the previous section allows for certain generalizations, among which also the SUSY extension of the BPS baby Skyrme model can be found. Later (in Section 7) we shall even find that the BPS baby Skyrme model allows for an $N=2$ extension. Concretely,  let us define the following lagrangians which generalize the quartic lagrangian of the previous subsection, 
\bea
\tilde L_\lambda &=& \int d^2 \theta \tilde{{\cal L}}_\lambda \equiv \int d^2\theta \lbrace D^\alpha\Phi D_\alpha \Phi^\dagger[D^2 \Phi D^2 \Phi^\dagger-\lambda D^\alpha D^\beta \Phi D_\alpha D_\beta\Phi^\dagger]\rbrace \\
\tilde L_\mu&=& \int d^2 \theta \tilde{{\cal L}}_\mu \equiv \int d^2\theta \lbrace D^\alpha\Phi D_\alpha \Phi[D^2 \Phi^\dagger D^2 \Phi^\dagger-\mu D^\alpha D^\beta \Phi^\dagger D_\alpha D_\beta\Phi^\dagger]\rbrace .
\eea
Here, $\lambda$ and $\mu$ are real parameters. 
In components, and for  the bosonic sector only, we get
\be
\tilde L_\lambda =(F\bar{F})^2(2-4\lambda)+(F\bar{F})(\partial_\mu u\partial^\mu\bar{u}) (2-8\lambda)-2\lambda(\partial_\mu u \partial^\mu \bar{u})^2
\ee
\bea
\tilde L_\mu&=& (F\bar{F})^2(2-4\mu)+\bar{F}^2(\partial_\mu u\partial^\mu u)(2-4\mu)-4\mu F^2\partial_\mu \bar{u}\partial^\mu\bar{u}-\\ \nonumber
&-&4\mu(\partial_\mu u \partial^\mu u)(\partial_\mu \bar{u} \partial^\mu \bar{u}) .
\eea
It follows that 
\bea
\mathfrak{Re}[\tilde L_\mu]&=&2(F\bar{F})^2(1-2\mu)+\bar{F}^2(\partial_\mu u\partial^\mu u)(1-4\mu)+\\ \nonumber
&+&F^2(\partial_\mu \bar{u}\partial^\mu \bar{u})(1-4\mu)-4\mu(\partial_\mu u \partial^\mu u)(\partial_\mu \bar{u} \partial^\mu \bar{u})
\eea
and, specifically for $\mu = 1/4$,
\be
\mathfrak{Re}[\tilde L_\mu]\vert_{\mu=\frac{1}{4}}=(F\bar{F})^2-(\partial_\mu u \partial^\mu u)(\partial_\mu \bar{u} \partial^\mu \bar{u})
\ee
A general linear combination of the two lagrangians is
\bea
\delta\mathfrak{Re}[\tilde L_\mu]\vert_{\mu=\frac{1}{4}}+\frac{\rho}{2}\tilde L_\lambda&=&(F\bar{F})^2(\delta+\rho-2\rho\lambda)+\rho(F\bar{F})(\partial_\mu u\partial^\mu\bar{u} )(1-4\lambda)-\\ \nonumber
&-&2\rho\lambda(\partial_\mu u \partial^\mu \bar{u})^2-\delta(\partial_\mu u \partial^\mu u)(\partial_\mu \bar{u} \partial^\mu \bar{u})
\eea
where $\delta$ and $\rho$ are real coefficients. Two choices for these parameters are of special interest, namely
\be
\tilde L^{(1)}_4 \equiv 
\left( \delta\mathfrak{Re}[\tilde L_\mu]+\frac{\rho}{2}\tilde L_\lambda\right)_{\mu=\frac{1}{4},\lambda=\frac{1}{4},\rho=-2,\delta =1}= (\partial_\mu u \partial^\mu \bar{u})^2-\vert \partial_\mu u\partial^\mu u\vert^2
\ee
and
\be
\tilde L^{(2)}_4 \equiv 
\left( \delta\mathfrak{Re}[\tilde L_\mu]+\frac{\rho}{2}\tilde L_\lambda\right)_{\mu=\frac{1}{4},\lambda=0,\rho=2,\delta=-1}=2(F\bar{F})(\partial_\mu u \partial^\mu \bar{u})+(F\bar{F})^2+\vert \partial_\mu u\partial^\mu u\vert^2 .
\ee
In a next step, we introduce, again, the target space area density $h(u, \bar u)$. This is done by multiplying the lagrangian densities in superspace by the corresponding superfield $h(U, U^\dagger)$ exactly like above, that is (where, again, we only consider the bosonic sector)
\be
\tilde L_\lambda = \int d^2 \theta \tilde{{\cal L}}_\lambda \quad \Rightarrow \quad L_\lambda = \int d^2 \theta h(U, U^\dagger) 
\tilde{{\cal L}}_\lambda = h(u,\bar u) \tilde L_\lambda
\ee
(and the same for $L_\mu$). The reason for this is that each superderivative $D_\alpha \Phi$ is linear in $\theta$ in the bosonic sector, and both ${\cal L}_\lambda$ and ${\cal L}_\mu$ are quadratic in $D_\alpha \Phi$ (i.e., quadratic in $\theta$ in the bosonic sector), therefore all superfields multiplying them only contribute with their $\theta =0$ component.  For the two quartic lagrangians $L^{(1)}_4 = h(u ,\bar u)\tilde L^{(1)}_4$ and
$L^{(2)}_4 = h(u ,\bar u)\tilde L^{(2)}_4$ we get 
\be \label{L(1)}
 L^{(1)}_4 = h(u , \bar u) \left( (\partial_\mu u \partial^\mu \bar{u})^2-\vert \partial_\mu u\partial^\mu u\vert^2 \right)
\ee
and
\be \label{L(2)}
 L^{(2)}_4 = h(u ,\bar u) \left( 
2(F\bar{F})(\partial_\mu u \partial^\mu \bar{u})+(F\bar{F})^2+\vert \partial_\mu u\partial^\mu u\vert^2 \right) .
\ee
The first expression (\ref{L(1)}) precisely coincides with the lagrangian (\ref{N1-L4}), therefore this choice of parameters just reproduces the $N=1$ extension of the previous section. In order to understand the significance of $L^{(2)}_4$, it is useful to add it to the quadratic lagrangian 
(\ref{l2}) of the previous section to obtain
\bea
L&=& L_2+ L^{(2)}_4=g(u,\bar{u})(\partial^\mu u \partial_\mu \bar{u}+F\bar{F}) + \nonumber \\
&+& h(u ,\bar u) \left(  2(F\bar{F})(\partial_\mu u \partial^\mu \bar{u})+(F\bar{F})^2+\vert \partial_\mu u\partial^\mu u\vert^2 \right) .
\eea
Now we solve for the auxiliary fields $F, \bar F$. On the one hand, we find the trivial solution $F=\bar F=0$ which leads to the lagrangian
\be
L = g(u,\bar{u})(\partial^\mu u \partial_\mu \bar{u}) + h(u , \bar u) +\vert \partial_\mu u\partial^\mu u\vert^2 .
\ee
This lagrangian contains higher than second powers of time derivatives, and we shall not consider it further in this paper. On the other hand, we find the nontrivial solution
\be
F\bar{F}=-\partial_\mu u \partial^\mu \bar{u}-\frac{g(u,\bar{u})}{2 h(u,\bar{u})}
\ee
and, after substituting back into the lagrangian,
\be
L=h(u,\bar{u})\lbrack  (\partial^\mu u\partial_\mu u)(\partial^\mu \bar{u}\partial_\mu \bar{u})-(\partial^\mu u\partial_\mu \bar{u})^2 \rbrack-\frac{g(u,\bar{u})^2}{4h(u,\bar{u})} .
\ee
For the choice $h=(1+u\bar u)^{-4}$, this is precisely the lagrangian of the BPS baby Skyrme model, where the quadratic term has disappeared, provided that we identify the potential with
\be
{\cal V}(u , \bar u) = \frac{g(u,\bar{u})^2}{4h(u,\bar{u})}.
\ee
As the quadratic term has disappeared, we are free to choose any function $g(u ,\bar u)$ we like and may, in this manner, produce the potentials we want. We emphasize that in this model the potential does not come from a superpotential but, instead, from the "target space metric" $g(u ,\bar u)$. Including a superpotential would result in a complicated fourth-order equation for the auxiliary field $F$, and the resulting lagragians would be completely different from the baby Skyrme model. We stop the discussion of the $N=1$ SUSY extension of the BPS baby Skyrme model at this point, because later we will find that this model allows, in fact, for an $N=2$ extension, such that also its BPS equations may be derived from $N=2$ SUSY (see Sections 7, 8).

\section{Gauged $N=1$ CP(1) baby Skyrme model}

In order to construct the gauged version of the $N=1$ $\mathbb{C}P^1$ baby-Skyrme model we need and extra superfield containing the gauge field $A_\mu$ and a Majorana fermion $\lambda_\alpha$ (in this case the photon and the photino field). This superfield, which we call $\Gamma_\alpha$, has the following decomposition,
\be
\Gamma_\alpha=i \theta^\beta(\gamma^\mu)_{\beta\alpha}A_\mu-2\theta^2\lambda_\alpha .
\ee 
In addition, we need the same complex superfield as above (constructed from two $N=1$ real superfields)
\be
U(x)=u(x)+\theta^\alpha \chi_\alpha(x)-\theta^2 F(x)
\ee
where $u(x)$ and $F(x)$ are complex fields and  $\chi_\alpha(x)$ is a Dirac fermion. Now it is easy to see that promoting the superderivative $D_\alpha$ to a covariant superderivative $\mathcal{D}_\alpha$,
\bea
\mathcal{D}^\alpha&=&D^\alpha+ie\Gamma^\alpha\\
\mathcal{D}_\alpha&=&D_\alpha-ie\Gamma_\alpha\\
\eea
and adding the Maxwell term, the model is automatically gauged. In close analogy to the ungauged case, the quadratic term for the gauged model is
\be
L_2^{g}=\int d^2\theta g(U^\dagger,U)\mathcal{D}^\alpha U^\dagger\mathcal{D}_\alpha U
\ee
and the bosonic (i.e., $\chi_\alpha = \lambda_\alpha =0$) sector results in
\be
L_2^{g}\vert_{bos} =g(\bar{u},u)(D^\mu \bar{u}D_\mu u+F\bar{F}) ,
\ee
where
\be
D_\mu u = \partial_\mu u +ieA_\mu u , \quad {D_\mu \bar u} = \partial_\mu \bar u -ieA_\mu \bar u .
\ee
Analogously, we find for the gauged quartic term
\begin{eqnarray}
{L}_{4a}^g&=&\int d^2\theta h(U^\dagger,U) \lbrack  \mathcal{D}^\alpha U \mathcal{D}_\alpha U + \mathcal{D}^\alpha U^\dagger \mathcal{D}_\alpha U^\dagger   \rbrack   \lbrack  \mathcal{D}^2 U \mathcal{D}^2 U +\mathcal{D}^2 U^\dagger \mathcal{D}^2 U^\dagger -\\
&-&  \frac{1}{4}(\mathcal{D}^\alpha \mathcal{D}^\beta U \mathcal{D}_\alpha \mathcal{D}_\beta U +\mathcal{D}^\alpha \mathcal{D}^\beta U^\dagger \mathcal{D}_\alpha \mathcal{D}_\beta U^\dagger ) \rbrack 
\end{eqnarray}
\begin{equation}
{L}_{4a}^g\vert_{bos}=h(\bar{u},u)\left((F^2+\bar{F}^2)^2-(D_\mu u)^2(D_\nu u )^2-(D_\mu \bar{u})^2(D_\nu \bar{u} )^2-2(D_\mu u)^2(D_\nu \bar{u} )^2\right)
\end{equation}
and, after again defining $A_1=U$ and $A_2=U^\dagger$,
\begin{equation}
{L}_{4b}^g=\sum_{ij}\int d^2\theta h(U^\dagger,U) \lbrack  \mathcal{D}^\alpha A^i \mathcal{D}_\alpha A^j   \rbrack   \lbrack  \mathcal{D}^2 V^i \mathcal{D}^2 V^j -  \frac{1}{4}(\mathcal{D}^\alpha \mathcal{D}^\beta A^i\mathcal{D}_\alpha \mathcal{D}_\beta A^j) \rbrack ,
\end{equation}
\begin{equation}
{L}_{4b}^g\vert_{bos}= h(\bar{u},u)\left((F^2+\bar{F}^2)^2-(D_\mu u)^2(D_\nu u )^2-(D_\mu \bar{u})^2(D_\nu \bar{u} )^2-2(D_\mu u D^\mu \bar{u})^2\right)
\end{equation}
and finally
\begin{equation}
{L}_{4}^g\vert_{bos}=-\frac{1}{2}({L}_{4a}^g\vert_{bos}-{L}_{4b}^g\vert_{bos})=-h(\bar{u},u)\left((D_\mu u D^\mu \bar{u})^2-(D_\mu u)^2(D_\nu \bar{u} )^2\right) .
\end{equation}
In addition, we need the Maxwell term which is generated in terms of the spinor superfield only,
\be
L_{M}= \frac{1}{8}\int d^2\theta \mathcal{D}^\beta\mathcal{D}^\alpha\Gamma_\beta \mathcal{D}^\gamma\mathcal{D}_\alpha\Gamma_\gamma .
\ee
Now we choose $g(\bar{u},u)=1/(1+\bar{u}u)^2$, $h(\bar{u},u)=1/(1+\bar{u}u)^4$. Putting all these terms together and eliminating the auxiliary fields we obtain in the bosonic sector
\begin{eqnarray}
L_{tot}^g&=&\frac{1}{(1+u\bar{u})^2} D_\mu u D^\mu \bar{u} - \frac{1}{(1+u\bar{u})^4}\lbrack (D_\mu u D^\mu \bar{u})^2-\\
&-&(D_\mu u D^\mu u)(D_\nu \bar{u} D^\nu \bar{u})  \rbrack- (1+u \bar{u})^2 {\cal U}_u {\cal U}_{\bar{u}}-\frac{1}{4}F_{\mu\nu}F^{\mu\nu}
\end{eqnarray}
where
\be
F_{\mu\nu}=\partial_\mu A_\nu-\partial_\nu A_\mu .
\ee
To summarize, we just find the gauged version of the baby Skyrme model, where partial derivatives are replaced by covariant derivatives, and a Maxwell term is included. This model is known to support soliton solutions \cite{GPS}. We remark that, exactly as in the ungauged case, within this SUSY extension it is not possible to eliminate the (gauged) quadratic, i.e., nonlinear sigma model, term without eliminating, at the same time, the potential. That is to say, we cannot construct the gauged BPS baby Skyrme model \cite{gaugedBPS-bS} within this SUSY extension. 
More general $N=1$ extensions which do allow to find the $N=1$ extension of the gauged BPS baby Skyrme model certainly will exist, like in the ungauged case (see Section 4.2). Here we shall consider, instead, directly the $N=2$ SUSY extension of the gauged BPS baby Skyrme model (Section 9), which turns out to exist, exactly as for the ungauged case.

\section{N=2 Supersymmetry in $2+1$ dimensions}

In this section we shall introduce our conventions for $N=2$ supersymmetry in $2+1$ dimensions. We have four independent Grassmann  variables, $\theta^\alpha$ and  $\bar{\theta}^{\dot{\alpha}}$, and the corresponding superderivatives 
\bea
D_\alpha&=&\frac{\partial}{\partial \theta^\alpha}+i\sigma_{\alpha\dot{\alpha}}^\mu\bar{\theta}^{\dot{\alpha}}\partial_\mu\\
\bar{D}_{\dot{\alpha}}&=&-\frac{\partial}{\partial \bar{\theta}^{\dot{\alpha}}}-i\theta^\alpha\sigma_{\alpha\dot{\alpha}}^\mu\partial_\mu .
\eea
With these definitions it is easy to check the following anticommutation relations,
\bea
\lbrace D_\alpha,  \bar{D}_{\dot{\alpha}} \rbrace&=&-2i\sigma_{\alpha\dot{\alpha}}^\mu\partial_\mu\\
\lbrace  D_\alpha, D_\beta  \rbrace&=&\lbrace   \bar{D}_{\dot{\alpha}} , \bar{D}_{\dot{\beta}}   \rbrace=0 .
\eea
The supersymmetric generators $Q$ and $\bar{Q}$ have the same structure as the superderivatives, up to a relative sign,
\bea
Q_\alpha&=&\frac{\partial}{\partial \theta^\alpha}-i\sigma_{\alpha\dot{\alpha}}^\mu\bar{\theta}^{\dot{\alpha}}\partial_\mu\\
\bar{Q}_{\dot{\alpha}}&=&\frac{\partial}{\partial \bar{\theta}^{\dot{\alpha}}}-i\theta^\alpha\sigma_{\alpha\dot{\alpha}}^\mu\partial_\mu ,
\eea
therefore the anticommutation relations are
\bea
\lbrace Q_\alpha,  \bar{Q}_{\dot{\alpha}} \rbrace&=&-2i\sigma_{\alpha\dot{\alpha}}^\mu\partial_\mu\\
\lbrace  Q_\alpha, Q_\beta  \rbrace&=&\lbrace   \bar{Q}_{\dot{\alpha}} , \bar{Q}_{\dot{\beta}}   \rbrace=0 ,
\eea
and the mixed anticommutators all vanish,
\be
\lbrace D_\alpha , Q_\beta  \rbrace=\lbrace D_\alpha ,\bar{Q}_{\dot{\beta}}  \rbrace=\lbrace\bar{D}_{\dot{\alpha}} ,Q_\beta  \rbrace=\lbrace \bar{D}_{\dot{\alpha}} ,\bar{Q}_{\dot{\beta}}  \rbrace=0 .
\ee
Now we introduce the superfields.  To construct our model, we will need only chiral and anti-chiral superfields satisfying the following constraints (for chiral and anti-chiral, respectively)
\bea
\bar{D}_{\dot{\alpha}}\Phi&=&0\\
D_\alpha\Phi^\dagger&=&0 .
\eea
 It is easy to solve the above constraints by introducing the chiral variables
 \be
 y^\mu=x^\mu+i\theta \sigma^\mu\bar{\theta}
 \ee
 (we assume dotted indices for variable with bar, and undotted without bar). These new variables satisfy the chiral constraint
 \be
 \bar{D}_{\dot{\alpha}}(x^\mu+i\theta \sigma^\mu\bar{\theta})=0,
 \ee
 therefore, by building superfields with this variable and expanding, the chiral constraint is automatically implemented. Concretely, for the chiral superfield
 \bea
 \Phi&=&u(x)+i\theta\sigma^\mu\bar{\theta}\partial_\mu u(x) +\frac{1}{4}\theta\theta\bar{\theta}\bar{\theta}\square u(x)+\sqrt{2}\theta\psi(x)-\\ \nonumber
 &-&\frac{i}{\sqrt{2}}\theta\theta\partial_\mu\psi(x)\sigma^\mu\bar{\theta}+\theta\theta F(x)
\label{chir}
 \eea
 and analogously for the anti-chiral superfield
  \bea
 \Phi^\dagger&=&\bar{u}(x)-i\theta\sigma^\mu\bar{\theta}\partial_\mu \bar{u}(x) +\frac{1}{4}\theta\theta\bar{\theta}\bar{\theta}\square \bar{u}(x)+\sqrt{2}\bar{\theta}\bar{\psi}(x)+\\ \nonumber
 &+&\frac{i}{\sqrt{2}}\bar{\theta}\bar{\theta}\theta\sigma^\mu\partial_\mu\bar{\psi}(x)+\bar{\theta}\bar{\theta}\bar{F}(x) .
\label{anti-chir}
 \eea

\section{The baby Skyrme model and N=2 supersymmetry}

In a first step, let us try to find an $N=2$ extension which produces the two kinetic terms $L_2$ and $L_4$,
\begin{equation}
L_2 + L_4 =\frac{\partial_\mu u\partial^\mu \bar{u}}{(1+\vert u \vert^2)^2}+ \frac{(\partial_\mu u)^2(\partial_\nu \bar{u})^2 - (\partial_\mu u\partial^\mu \bar{u})^2}{(1+\vert u \vert^2)^4} .
\label{fad lag}
\end{equation}
In order to generate the quadratic term, we need only a D-term involving a Kahler potential (this is just the $N=2$  CP(1) $\sigma$-model),
with the lagrangian density
\begin{equation} \label{n2-sigma-lag}
L_2=\frac{1}{16}\int  d^2\theta  d^2\bar{\theta}\;  \mbox{ln} (1+\Phi\Phi^\dagger)
\end{equation}
where $\Phi$ is a N=1 chiral superfield in $(2+1)$ dimensions and ${\Phi}^\dagger $ the respective antichiral superfield. Taking into account that 
$K(\Phi ,\Phi^\dagger )= \mbox{ln} (1+\Phi\Phi^\dagger)$ is a Kaehler potential with Kaehler metric
\be
g(u,\bar{u})=g_{\bar{u} u}=\partial_u \partial_{\bar u} K(u ,\bar u) =\frac{1}{(1+\bar{u}u)^2}
\ee
the only non-zero Christoffel symbols are
\bea
\Gamma^u_{u}&=&g^{u\bar{u}}\partial_u g(u,\bar{u})\\
\Gamma^{\bar{u}}_{\bar{u}\bar{u}}&=&g^{\bar{u}u}\partial_{\bar{u}} g_{\bar{u} u}
\eea
or, explicitly,
\bea
\Gamma^u_{uu}&=&\frac{-2\bar{u}}{1+u\bar{u}}\\
\Gamma^{\bar{u}}_{\bar{u}\bar{u}}&=&\frac{-2u}{1+u\bar{u}} .
\eea
The lagrangian can be written in components as
\be
L_2 = g(u,\bar{u}) \lbrack \partial^\mu u\partial_\mu \bar{u}-\frac{i}{2}\psi\sigma^\mu \mathcal{D}^\mu \bar{\psi} +\frac{i}{2}\mathcal{D}^\mu\psi\sigma_\mu\bar{\psi}+ F\bar{F}\rbrack
+ \frac{1}{4}\mathcal{R}_{u\bar{u} u\bar{u}}(\psi\psi)(\bar{\psi}\bar{\psi})
\ee
where
\bea
 g(u,\bar{u}) &=&\frac{1}{(1+u\bar{u})^2}\\
 \mathcal{R}_{u\bar{u} u\bar{u}}&=&-\frac{2}{(1+u\bar{u})^4}\\
 \mathcal{D}_\mu\psi_\alpha&\equiv& \left( \partial_\mu -\frac{2\bar{u}}{1+u\bar{u}} \partial_\mu u \right) \psi_\alpha\\
  \mathcal{D}_\mu\psi^{\dagger\dot{\alpha}}&\equiv& \left( \partial_\mu -\frac{2u}{1+u\bar{u}} \partial_\mu \bar{u}\right) \psi^{\dagger\dot{\alpha}} .
\eea
In a next step, we have to generate the $N=2$ supersymmetric version of the quartic terms in (\ref{fad lag}). We might choose a supersymmetric lagrangian starting from a superfield quartic in superderivatives and depending on both chiral and anti-chiral superfields. Let ${\mathcal{\tilde L}}_4$ be this quartic superfield,
\be
\mathcal{\tilde{L}}_4=\frac{1}{16}D^\alpha\Phi D_\alpha\Phi \bar{D}^{\dot{\beta}}\Phi^\dagger  \bar{D}_{\dot{\beta}}\Phi^\dagger
\ee
then after integration in the Grassmann variables we get for the bosonic sector
\be
{\tilde{L}}_{4,bos}= (\partial^\mu u)^2(\partial^\nu \bar{u})^2 + 2\bar{F} F \partial^\mu u \cdot \partial_\mu \bar{u} + (\bar{F}F)^2 .
\ee
Right now, this quartic lagragian is still quite different from the quartic part of (\ref{fad lag}). The first observation in that we can multiply this lagrangian by a prefactor depending on the superfields. Let this prefactor be  $h(\Phi,\Phi^\dagger)$, then the new superfield has the following form,
\be
\mathcal{L}_4=\frac{1}{16}h(\Phi,\Phi^\dagger)D^\alpha\Phi D_\alpha\Phi \bar{D}^{\dot{\beta}}\Phi^\dagger  \bar{D}_{\dot{\beta}}\Phi^\dagger
\ee
and 
after the $\theta-$integration the bosonic sector of the corresponding lagrangian is 
\be
{L}_{4,bos} = h(u,\bar{u})\lbrack(\partial^\mu u)^2(\partial^\nu \bar{u})^2 + 2\bar{F} F \partial^\mu u \cdot \partial_\mu \bar{u} + (\bar{F}F)^2\rbrack .
\ee
The reason for this result is that each superderivative $D^\alpha \Phi$ is at least linear in $\theta$ or $\bar \theta$ in the bosonic sector, and the above superfield contains four powers of $D^\alpha \Phi$'s. Therefore, only the $\theta$-independent part of the prefactor contributes to the bosonic sector.

Adding the bosonic sector of the quadratic lagrangian to the above quartic bosonic lagrangian we get
\bea
{L}_{T,bos}&=&g(u ,\bar u) \lbrack \partial^\mu u\partial_\mu \bar{u}+ F\bar{F}\rbrack+\\ \nonumber
&+&h(u,\bar u)\lbrack(\partial^\mu u)^2(\partial^\nu \bar{u})^2 + 2\bar{F} F \partial^\mu u \cdot \partial_\mu \bar{u} + (\bar{F}F)^2\rbrack
\eea
Finally solving the algebraic equation of motion for $F\bar{F}$:
\be
F\bar{F}= - \partial_\mu u\partial^\mu \bar{u}-\frac{g(u,\bar{u})}{2h(u,\bar{u})}
\ee
we get 
\be
{L}_{T,bos} = h(u ,\bar u)\lbrack (\partial^\mu u\partial_\mu u)(\partial^\mu \bar{u}\partial_\mu \bar{u})-(\partial^\mu u\partial_\mu \bar{u})^2  \rbrack - \frac{g (u ,\bar u)^2}{4h(u ,\bar u)}.
\ee
For our special case with
 $g(u,\bar{u})=1/(1+u\bar{u})^2$, $h(u,\bar{u})=1/(1+u\bar{u})^4$, this turns into
\be
{L}_{T,bos} = \frac{1}{(1+u \bar{u})^4}\lbrack (\partial^\mu u\partial_\mu u)(\partial^\mu \bar{u}\partial_\mu \bar{u})-(\partial^\mu u\partial_\mu \bar{u})^2  \rbrack - \frac{1}{4},
\ee
so we apparently find a constant "potential" ${\cal V}=(1/4)$. 
The important observation here is that after the substitution of the auxiliary field $F$ by its on-shell value, the quadratic, nonlinear sigma model term has completely disappeared from the above bosonic lagrangian, for arbitrary choices of $g$ and $h$. There is, therefore, no more reason to restrict the Kaehler metric $g$ and the corresponding Kaehler potential to their CP(1) form.    
Then, choosing $h(u,\bar{u})=1/(1+u\bar{u})^4$ (which we maintain, because we want the standard quartic term of the baby Skyrme model), and a general Kaehler manifold (different from CP(1)) with metric $g(u,\bar{u})$ we have in the bosonic sector
\bea
{L}_{T,bos}&=&\frac{1}{(1+u \bar{u})^4}\lbrack (\partial^\mu u\partial_\mu u)(\partial^\mu \bar{u}\partial_\mu \bar{u})-(\partial^\mu u\partial_\mu \bar{u})^2  \rbrack-\\
&-&\frac{g(u,\bar{u})^2}{4}(1+u\bar{u})^4 . \nonumber
\eea
This is precisely the lagrangian of the BPS baby Skyrme model with the potential term
\be 
{\cal V}(u,\bar u) = \frac{g(u,\bar{u})^2}{4}(1+u\bar{u})^4
\ee
where $g$ is a Kaehler metric. The potential in the $N=2$ extension is, therefore, induced by the Kaehler potential of the nonlinear sigma-model type lagrangian (\ref{n2-sigma-lag}), and {\em not} by a superpotential term. The addition of a superpotential is, in fact, forbidden in the sense that it would transform the algebraic field equation of the auxiliary field $F$ into a fourth order equation with complicated roots of $u$ and $\partial_\mu u$ as solutions. The resulting lagrangian would then be completely different from the baby Skyrme lagrangian.

To summarize, we found the $N=2$ supersymmetric extension of the restricted baby Skyrme model. Let us give some concrete examples. For  the following family of potentials ${\cal V}(u,\bar{u})$ depending on the parameter $s$,
\be
{\cal V}(u,\bar{u})=\left(\frac{u\bar{u}}{1+u\bar{u}}\right)^s
\ee
the corresponding Kaehler metrics generating these potentials are
\be
g(u,\bar{u})=2\frac{(u\bar{u})^\frac{s}{2}}{(1+u\bar{u})^\frac{s+4}{2}}.
\ee
Integrating this metric we obtain the Kaehler potential, which at the superfield level is
\be
K(\Phi,\Phi^\dagger)=\frac{8(\Phi\Phi^\dagger)^\frac{s+2}{2}}{(2+s)^2} {}_2F_1[\frac{2+s}{2},\frac{2+s}{2},\frac{4+s}{2},-\Phi\Phi^\dagger] ,
\ee
for example,
\bea
s=1,&\quad& K(\Phi,\Phi^\dagger )=\mbox{arcsinh} \, (\sqrt{\Phi\Phi^\dagger})-\sqrt{\frac{\Phi\Phi^\dagger}{1+\Phi\Phi^\dagger}}\\
s=2,&\quad& K(\Phi,\Phi^\dagger )=\frac{1}{1+\Phi\Phi^\dagger}+\mbox{ln}\, (1+\Phi\Phi^\dagger ) .
\eea
Reintroducing the coupling constant $\mu$ of the potential terms, we get the bosonic lagrangians
\bea
s=1,&\quad& {L}_T^1=\frac{1}{(1+u \bar{u})^4}\lbrack (\partial^\mu u\partial_\mu u)(\partial^\mu \bar{u}\partial_\mu \bar{u})-(\partial^\mu u\partial_\mu \bar{u})^2  \rbrack-\\ \nonumber
&-&2\mu^2\left(\frac{u\bar{u}}{1+u\bar{u}}\right)\\ 
s=2,&\quad& {L}_T^2=\frac{1}{(1+u \bar{u})^4}\lbrack (\partial^\mu u\partial_\mu u)(\partial^\mu \bar{u}\partial_\mu \bar{u})-(\partial^\mu u\partial_\mu \bar{u})^2  \rbrack-\\ \nonumber
&-&2\mu^2\left(\frac{u\bar{u}}{1+u\bar{u}}\right)^2\\ \nonumber
...
\eea
We remark that
the parameter $\mu$ is introduced in the D-term generated by the Kaehler potential, hence it is present in the fermionic sector of this term.

\section{Bogomol'nyi equation}
The BPS baby Skyrme model is well-known to support BPS solitons, that is, solitons which saturate the Bogomol'nyi bound and obey the corresponding first order BPS equation. In addition, we just found that the model admits an $N=2$ SUSY extension, so the natural question arises whether these BPS solitons may be recovered as one-half BPS states of the supersymmetrically extended theory. SUSY BPS states are characterized by the fact that they are annihilated by some of the SUSY charges or, in the case of classical BPS solutions, that some SUSY charges (SUSY transformations) are zero when evaluated for the BPS states. We, therefore, need the $N=2$ SUSY transformations in a first step. 
More concretely, a SUSY BPS solution has the fermionic components of the basic superfield $\Phi$ equal to zero, and only the scalar field $u$ and the auxiliary field $F$ are nontrivial. Further, the SUSY transformation of both $u$ and $F$ is proportional to a fermion and therefore trivially zero for a BPS state. The only nontrivial conditions, thus, come from the SUSY transformations of the spinors.
The $N=2$ transformations of the spinors have the following form
\bea
\delta\psi_\beta&=& - i\partial_{\beta\dot{\alpha}} u \bar{\epsilon}^{\dot{\alpha}}+F\epsilon_\beta\\
\delta\bar{\psi}^{\dot{\beta}}&=& i\partial^{\dot{\beta}\alpha} \bar{u} \epsilon_\alpha+\bar{F}\bar{\epsilon}^{\dot{\beta}}
\eea
where $\epsilon_\alpha$ and $\bar \epsilon^{\dot \alpha}$ are the Grassmann-valued SUSY transformation parameters.
For static (time-independent) fields we find in components
\bea
\delta\psi_1\vert_{static}&=& \partial_1 u \bar{\epsilon}^{\dot{1}} - \partial_2 u \bar{\epsilon}^{\dot{2}} +F\epsilon_1\\
\delta\psi_2\vert_{static}&=& - \partial_1 u \bar{\epsilon}^{\dot{2}} - \partial_2 u \bar{\epsilon}^{\dot{1}} +F\epsilon_2\\
\delta\bar\psi^{\dot 1}\vert_{static}&=& \partial_1\bar{u} \epsilon_1 - \partial_2 \bar{u} \epsilon_2 +\bar{F}\bar{\epsilon}^{\dot{1}}\\
\delta\bar\psi^{\dot 2}\vert_{static}&=& - \partial_1 \bar{u} \epsilon_2 - \partial_1 \bar{u} \epsilon_1 +\bar{F}\bar{\epsilon}^{\dot{2}}
\eea
or
\be 
\delta \vec \psi = M \vec \epsilon
\ee
where $\vec \psi = (\psi_1 ,\psi_2 , \bar\psi^{\dot 1} , \bar\psi^{\dot 2} )^t$, $\vec \epsilon = ( \epsilon_1 ,\epsilon_2 ,\bar \epsilon^{\dot 1}, \bar \epsilon^{\dot 2})^t$, and $M$ is the matrix 
\be
 M = \left(\begin{array}{cccc}
 \partial_1 u & - \partial_2 u & F&0 \\
- \partial_2 u & - \partial_1 u & 0 & F\\
\bar{F} & 0 & \partial_1 \bar{u} & - \partial_1\bar{u}\\
0 & \bar{F} & - \partial_2\bar{u} & - \partial_1\bar{u} \end{array} \right) .
\ee
The condition that some (linear combinations of the) SUSY transformations $\delta \psi$ are zero is equivalent to the condition $\mbox{det } M=0$, therefore we now need the eigenvalues of $M$. These eigenvalue may be calculated to be $(\lambda_+ ,-\lambda_+ ,\lambda_- ,- \lambda_-)$, where
\be
\lambda_\pm^2 =  - \partial_i u \partial^i\bar{u}\pm\sqrt{(\partial_i u \partial^i \bar{u})^2-(\partial_i u)^2(\partial_j \bar{u})^2} -F\bar F,
\ee
and the determinant is
\be
\mbox{det} M=(\partial_i u)^2(\partial_j \bar{u})^2 + 2\partial_i u \partial^i\bar{u} F\bar{F}+(F\bar{F})^2 .
\ee
The condition $\mbox{det } M=0$ therefore leads either to $\lambda_+^2 =0$, that is,
\be
F\bar{F}= - \partial_i u \partial^i\bar{u} + \sqrt{(\partial_i u \partial^i \bar{u})^2-(\partial_i u)^2(\partial_j \bar{u})^2}
\ee
or to $\lambda_-^2 =0$, that is,
\be
F\bar{F}= - \partial_i u \partial^i\bar{u} - \sqrt{(\partial_i u \partial^i \bar{u})^2-(\partial_i u)^2(\partial_j \bar{u})^2},
\ee
corresponding to soliton and antisoliton, respectively.
As the eigenvalues come in pairs, each condition has multiplicity two, and possible BPS solutions are, therefore, always one-half BPS states (they leave invariant one-half of the supersymmetries). We remark that the discussion up to now has been completely general, and the above equations are therefore the {\em completely general} one-half BPS equations for any $N=2$ supersymmetric field theory constructed from a chiral superfield. Specific models are characterized by the specific field equations for the auxiliary field $F$. 

Concretely, for the $N=2$ BPS baby Skyrme model, the equation of motion for $F\bar{F}$ in the static regime is
\be \label{N2-F-eq}
F\bar{F}= - \partial_i u\partial^i\bar{u}-\frac{g(u,\bar{u})}{2h(u,\bar{u})} ,
\ee
and we obtain the BPS equations
\be
\mp \sqrt{ (\partial_i u \partial^i \bar{u})^2-(\partial_i u)^2(\partial_j \bar{u})^2 }=\frac{g(u,\bar{u})}{2h(u,\bar{u})} .
\ee
In order to demonstrate that this is, indeed, precisely the BPS equation of the BPS baby Skyrme model, we use 
\be
(\partial_i u \partial^i \bar{u})^2-(\partial_i u)^2(\partial_j \bar{u})^2 = (i\epsilon_{jk}u_j \bar u_k)^2
\ee
and the expression for the topological charge density $q\equiv K^0/2$ where
\be
q(x) =\frac{2i\epsilon_{jk}u_j\bar u_k}{(1+u\bar u)^2} , \quad \int d^2 x q (x) = 4\pi k.
\ee
The normalization of $q$ is useful because then $q$ is just the pullback (under the map defined by $u$) of the area two-form on the target space unit sphere (the area of the unit sphere is $4\pi$). Using this expression, and $h=(1+u\bar u)^{-4}$, we get for the BPS equation
\be \label{BPS-bS-BPS-eq}
q(x) = \pm g(u,\bar u) (1+u\bar u)^2 = \pm 2 \sqrt{{\cal V}(u ,\bar u)}
\ee
where  
\be
{\cal V}(u,\bar{u})=\frac{1}{4}g(u,\bar{u})^2 (1+u\bar u)^4 .
\ee
This is precisely the BPS equation of the BPS baby Skyrme model, see e.g. \cite{restr-bS}, \cite{Sp1} (in those papers the r.h.s. of Eq. (\ref{BPS-bS-BPS-eq}) reads $\pm \sqrt{2{\cal V}}$, because there the potential shows up in the lagrangian like $-(\mu^2/2){\cal V}$, whereas it appears without the factor $1/2$ in the present paper).

Remark: it might appear that the on-shell value (\ref{N2-F-eq}) for $F\bar F$ is negative, which would be contradictory. Here we want to show that, at least for field configurations which are sufficiently close to the BPS bound, this is not the case. Indeed, from (\ref{N2-F-eq}) we easily derive
\be
\frac{2F\bar F}{(1+u\bar u)^2} = \frac{2\nabla u \cdot \nabla \bar u}{(1+u\bar u)^2 } - 2 \sqrt{{\cal V}}
\ee
and, using the BPS equation (\ref{BPS-bS-BPS-eq}), 
\be
\frac{2F\bar F}{(1+u\bar u)^2} = \frac{2\nabla u \cdot \nabla \bar u}{(1+u\bar u)^2 } \pm q(x) \ge 0.
\ee

\section{N=2 SUSY gauged  Skyrme model in 2+1 dimensions}

Recently it has been found that the gauged BPS baby Skyrme model still has a BPS bound and supports soliton solutions saturating this bound \cite{gaugedBPS-bS}, so it is natural to attempt an $N=2$ SUSY extension for this case, as well. For this purpose, we need the formalism for $N=2$ supersymmetric gauge fields, concretely for abelian gauge fields (Maxwell electrodynamics). For the gauged version of the Kaehler potential term (i.e., the quadratic kinetic term), we use the well-known fact that the combination of superfields $\Phi^\dagger e^V \Phi$ is gauge invariant, where $V$ is the real vector multiplet with components (in the Wess-Zumino gauge)
\be
V=-\theta \sigma^\mu \bar{\theta}A_\mu+i\theta\theta\bar{\theta}\bar{\lambda}-i\bar{\theta}\bar{\theta}\theta\lambda-\frac{1}{2}\theta\theta\bar{\theta}\bar{\theta}D+\theta\gamma^5\bar{\theta}\sigma .
\ee
Here $D$ and $\sigma$ are real fields. Again we need the chiral and antichiral superfields (\ref{chir}), (\ref{anti-chir}), which are $N=2$ supersymmetric by construction. The gauged quadratic term may now be constructed starting from
\be
{L}^g_2=\int d^2\theta d^2\bar{\theta} K(\Phi^\dagger e^V \Phi)
\ee
where $K$ is a generalized Kaehler potential. Integrating we obtain the lagrangian
\bea
{L}^g_2&=& g_{u\bar{u}} \left(  D^\mu u D_\mu \bar{u}-\frac{i}{2}\psi\sigma^\mu \mathcal{D}^\mu \bar{\psi} +\frac{i}{2}\mathcal{D}^\mu\psi\sigma_\mu\bar{\psi}+ F\bar{F}\right) \nonumber \\ 
&+&\frac{1}{4}\mathcal{R}_{u\bar{u} u\bar{u}}(\psi\psi)(\bar{\psi}\bar{\psi})+\sigma^2\bar{u} u g_{u\bar{u}}+ \left( u\frac{\partial K}{\partial u}+\bar{u}\frac{\partial K}{\partial \bar{u}}\right) D- \nonumber \\
&-&ig_{u\bar{u}}(u\lambda\psi+\bar{u}\bar{\lambda}\bar{\psi}) .
\label{L-2-g}
\eea

Here, $D^\mu$ is the standard covariant derivative, and $\mathcal{D}^\mu$ is the covariant derivative on spinors,
\be
\mathcal{D}_\mu\psi=\partial_\mu \psi+(\partial_\mu u)\Gamma_{uu}^u\psi+ieA_\mu\psi .
\ee
Further,  $g_{u\bar{u}}$  is the Kaehler metric.

In a next step, we have to covariantize the quartic term. This is easily done by introducing the spinor gauge superfields defined by
\bea
\Gamma_\alpha&=&D_\alpha V\\
\bar{\Gamma}^{\dot{\alpha}}&=&\bar{D}^{\dot{\alpha}} V
\eea
and changing the superderivatives to the covariant superderivatives, $\tilde{D}_\alpha$ and $\tilde{\bar{D}}^{\dot{\alpha}}$,
\bea
\tilde{D}_\alpha&=&D_\alpha +\Gamma_\alpha\\
\tilde{\bar{D}}^{\dot{\alpha}}&=&\bar{D}^{\dot{\alpha}} +\tilde{\bar{\Gamma}}^{\dot{\alpha}} ,
\eea
hence $L_{4}^g$ is the $\theta^2 \bar\theta^2$ component of the superfield
\be
\mathcal{L}_4^g=\frac{1}{16}h(\Phi,\Phi^\dagger)\tilde{D}^\alpha\Phi \tilde{D}_\alpha\Phi \bar{\tilde{D}}^{\dot{\beta}}\Phi^\dagger  \bar{\tilde{D}}_{\dot{\beta}}\Phi^\dagger .
\ee
The bosonic part of this lagrangian reads
in components
\be
{L}_{4,bos}^g=h(u,\bar u)\lbrack(D^\mu u)^2(D^\nu \bar{u})^2 + 2\bar{F} F D^\mu u \cdot D_\mu \bar{u} + F^{\dagger 2}F^2\rbrack+O(\sigma^2) .
\ee
Finally, we need the $N=2$ extension of the Maxwell lagrangian, which is constructed from the superfields
\bea
W_\alpha&=&-\frac{1}{4}\bar{D}\bar{D} D_\alpha V\\
\bar{W}_{\dot{\alpha}}&=&-\frac{1}{2}DD\bar{D}_{\dot{\alpha}}V .
\eea
The corresponding Maxwell lagrangian then is
\be \label{L-M-N2-1}
L_{M}=\frac{1}{4}(W^\alpha W_\alpha\vert_{\theta\theta}+\bar{W}_{\dot{\alpha}}\bar{W}^{\dot{\alpha}}\vert_{\bar{\theta}\bar{\theta}})
\ee
or
\be \label{L-M-N2-2}
L_{M}=-\frac{1}{4}F_{\mu\nu}F^{\mu\nu}+\frac{1}{2}D^2+\frac{1}{2}\partial^\mu\sigma\partial_\mu\sigma-i\lambda\gamma^\mu\partial_\mu\bar{\lambda} .
\ee
The complete lagrangian in the bosonic sector, therefore, reads
\bea
{L}_b^g&=& g_{u \bar{u}} \left(  D^\mu u D_\mu \bar{u}+ F\bar{F}+(u\frac{\partial K}{\partial u}+\bar{u}\frac{\partial K}{\partial \bar{u}})D\right) +\\ \nonumber
&+&h(u,\bar{u})\left( (D^\mu u)^2(D^\nu \bar{u})^2 + 2\bar{F} F D^\mu u  D_\mu \bar{u} + F^{\dagger 2}F^2\right) -\\ \nonumber
&-&\frac{1}{4}F_{\mu\nu}F^{\mu\nu}+\frac{1}{2}D^2 +O(\sigma^2) .
\eea
The real scalar field $\sigma$ appears at least quadratically, therefore, the trivial vacuum configuration $\sigma =0$ always is a solution.
We eliminate $\sigma$ using this trivial solution. Further,
the (algebraic) field equations for the auxiliary fields $F$ and $D$ are solved by
\be
F\bar{F}= - D^\mu u D_\mu \bar u-\frac{g(u,\bar{u})}{2 h(u,\bar{u})}
\ee
\be
D=-\left( u\frac{\partial K}{\partial u}+\bar{u}\frac{\partial K}{\partial \bar{u}}\right)
\ee
and,
using them (and $\sigma =0$), the complete bosonic lagrangian finally reads 
\bea
{L}_b^g&=&h(u,\bar{u})\left( (D^\mu u)^2(D^\nu \bar{u})^2-(D^\mu u D_\mu \bar{u})^2 \right) -\frac{1}{4}F_{\mu\nu}F^{\mu\nu}\\ \nonumber
&-&\frac{g(u,\bar{u})^2}{4 h(u,\bar{u})}-\frac{1}{2}\left( u\frac{\partial K}{\partial u}+\bar{u}\frac{\partial K}{\partial \bar{u}}\right)^2 .
\eea
We, therefore, found a bosonic lagrangian where the quadratic, sigma-model type contribution has disappeared, again, the quartic Skyrme term is covariantized, a Maxwell term has been created, and, finally, a potential has been produced by the two auxiliary fields $F$ and $D$, which explicitly reads
\be
\mathcal{V}(u,\bar{u})=\frac{g(u,\bar{u})^2}{4 h(u,\bar{u})}+\frac{1}{2}\left( u\frac{\partial K}{\partial u}+\bar{u}\frac{\partial K}{\partial \bar{u}}\right)^2 
\ee
or
\be
\mathcal{V}(u,\bar{u})=\frac{1}{4 h(u,\bar{u})}\left( \frac{\partial^2 K}{\partial u \partial {\bar u}} \right)^2 +\frac{1}{2}\left( u\frac{\partial K}{\partial u}+\bar{u}\frac{\partial K}{\partial \bar{u}}\right)^2 .
\ee
For later use we now assume that $h(u ,\bar u) = h(u\bar u)$ and $K(u,\bar u)=K(u\bar u)$, and define $K' \equiv \partial_{u\bar u} K$, then
\be
\mathcal{V}(u \bar{u})=\frac{1}{4 h(u \bar{u})}\left( K' + u\bar u K'' \right)^2 + 2 \left( u\bar u K' \right)^2
\ee
or
\be
\mathcal{V}(u \bar{u})=\frac{1}{4 h(u \bar{u})}{\cal W}'^2  + 2 {\cal W}^2 \; , \quad {\cal W} \equiv u\bar u K'.
\ee

\subsection{Bogomol'nyi equations}
In a next step, we want to recover the BPS equations of the gauged BPS baby Skyrme model as one-half BPS states of the $N=2$ supersymmetrically extended theory, as in the ungauged model. For this purpose,
 we need the supersymmetric transformations of the multiplet (for fermions)
\bea
\delta\lambda_\alpha&=&-\eta_\alpha D-\frac{1}{2}\epsilon^{\mu\nu\lambda}F_{\mu\nu}(\gamma_\lambda)_\alpha^\beta\eta_\beta-i(\gamma^\mu)_\alpha^\beta\partial_\mu \sigma\eta_\beta\\
\delta\bar{\lambda}^{\dot{\alpha}}&=&-\bar{\eta}^{\dot{\alpha}}D-\frac{1}{2}\epsilon^{\mu\nu\lambda}F_{\mu\nu}(\gamma_\lambda)^{\dot{\alpha}}_{\dot{\beta}}\bar{\eta}^{\dot{\beta}}+i(\gamma^\mu)^{\dot{\alpha}}_{\dot{\beta}}\partial_\mu \sigma \bar{\eta}^{\dot{\beta}}\\
\delta\psi_\beta&=& - iD_{\beta\dot{\alpha}} u \bar{\epsilon}^{\dot{\alpha}}+F\epsilon_\beta+i\eta_\alpha\sigma u\\
\delta\bar{\psi}^{\dot{\beta}}&=& iD^{\dot{\beta}\alpha} \bar{u} \epsilon_\alpha+\bar{F}\bar{\epsilon}^{\dot{\beta}}-i\bar{\eta}_{\dot{\alpha}}\sigma \bar{u} .
\eea
We, again, restrict to the trivial solution
$\sigma=0$ for the $\sigma$ field to obtain
\bea
\delta\lambda_\alpha&=&-\eta_\alpha D-\frac{1}{2}\epsilon^{\mu\nu\lambda}F_{\mu\nu}(\gamma_\lambda)_\alpha^\beta\eta_\beta\\
\delta\bar{\lambda}^{\dot{\alpha}}&=&-\bar{\eta}^{\dot{\alpha}}D-\frac{1}{2}\epsilon^{\mu\nu\lambda}F_{\mu\nu}(\gamma_\lambda)^{\dot{\alpha}}_{\dot{\beta}}\bar{\eta}^{\dot{\beta}}\\
\delta\psi_\beta&=& - iD_{\beta\dot{\alpha}} u \bar{\epsilon}^{\dot{\alpha}}+F\epsilon_\beta\\
\delta\bar{\psi}^{\dot{\beta}}&=& iD^{\dot{\beta}\alpha} \bar{u} \epsilon_\alpha+\bar{F}\bar{\epsilon}^{\dot{\beta}} .
\eea
Now
we are ready to repeat the strategy of the ungauged model. That is to say, we have to calculate the matrices of the susy transformations of both spinors, take the determinants (or their eigenvalues) and extract the Bogomol'nyi equations. In the last step we then have to take into account the on-shell values  of the auxiliary fields. The matrices of the SUSY transformations for static fields are
\[ M_\psi\vert_s = \left(\begin{array}{cccc}
 D_1 u & - D_2 u & F&0 \\
- D_2 u & - D_1 u & 0 & F\\
\bar{F} & 0 & D_1 \bar{u} & - D_2\bar{u}\\
0 & \bar{F} & -D_2\bar{u} & -D_1\bar{u} \end{array} \right)\] 
and (where we also assume $A_0 =0$)
\[ M_\lambda\vert_s = \left(\begin{array}{cccc}
-D  & \frac{i}{2}\epsilon^{ij}F_{ij} & 0&0 \\
-\frac{i}{2}\epsilon^{ij}F_{ij}  & -D & 0 & 0\\
0 & 0 &-D &\frac{i}{2}\epsilon^{ij}F_{ij} \\
0 &0&-\frac{i}{2}\epsilon^{ij}F_{ij} &-D \end{array} \right) , \] 
and from
 $\mbox{det}( M_\psi\vert_s)=0$ and $\mbox{det}(M_\lambda\vert_s )=0$ we obtain the general BPS equations
\bea
F\bar{F}&=& - D_iu D_i \bar{u}\pm\sqrt{(D_iu D_i \bar{u})^2-(D_i u)^2(D_j \bar{u})^2} \label{B1}\\
 D&=&\pm\epsilon^{ij}F_{ij} .
 \label{B2}
\eea
We emphasize that, again, these are the completely general BPS equations for a general $N=2$ chiral superfield coupled to an $N=2$ extended  abelian gauge field. Specific models result from specific solutions for the auxiliary fields $F$ and $D$.

Concretely, for the gauged BPS baby Skyrme model we get
\bea
\frac{g}{2h}&=&\pm\sqrt{(D_iu D_i \bar{u})^2-(D_i u)^2(D_j \bar{u})^2} \label{B1a}\\
 \left( u\frac{\partial K}{\partial u}+\bar{u}\frac{\partial K}{\partial \bar{u}}\right) &=&\pm\epsilon^{ij}F_{ij} .
 \label{B2a}
\eea
For a comparison to known results it is useful to simplify the square root in the first equation,
\be
(D_iu D_i \bar{u})^2-(D_i u)^2(D_j \bar{u})^2 = (i\epsilon_{jk}D_juD_k \bar u)^2
\ee
and
\be
i\epsilon_{jk} D_j u D_k \bar u = i\epsilon_{jk} u_j \bar u_k + e \epsilon_{jk} A_k \partial_j (u\bar u) ,
\ee 
then we  get 
\bea
\frac{{\cal W}'}{2h}&=&\pm \left( i\epsilon_{jk} u_j \bar u_k + e\epsilon_{jk} A_k \partial_j (u\bar u) \right)   \\
 2{\cal W} &=&\pm\epsilon^{ij}F_{ij} 
 \label{B2b0}
\eea
where we also assumed $K=K(u\bar u)$, as above. Introducing now the topological charge density $q$ and its "covariant" version $Q$,
\be
Q = \frac{i\epsilon_{jk} D_j u D_k \bar u}{(1+u\bar u)^2 } = q + \frac{e}{(1+u\bar u )^2} \epsilon_{jk} A_k \partial_j (u\bar u)
\ee
and using the explicit expression $h=(1+u\bar u)^{-4}$, we finally get the BPS equations
\bea
\frac{(1+u \bar u)^2}{2}{\cal W}'&=&\pm Q \label{B1b1} \\
 {\cal W} &=&\pm B
 \label{B2b1}
\eea
where $B$ is the magnetic field, $B=\epsilon_{ij} \partial_i A_j = F_{12}$. For a direct comparison with the results of \cite{gaugedBPS-bS} we should take into account that in that paper the potential ${\cal V}$ and the "superpotential" ${\cal W}$ were treated as functions of $n_3$ instead of $u\bar u$, where
\be
n_3 = \frac{1-u\bar u}{1+u\bar u} \quad \Rightarrow \quad \partial_{u\bar u} = -\frac{2}{(1+u\bar u)^2} \partial_{n_3}
\ee
which leads to the BPS equations
\bea
{\cal W}_{n_3} &=&\mp Q \label{B1b} \\
 {\cal W} &=&\pm B.
 \label{B2b}
\eea
These are precisely the BPS equations of Ref. \cite{gaugedBPS-bS}, after the corresponding coupling constants have been reintroduced. Finally, for the relation between ${\cal W}$ and ${\cal V}$ we get
\be
{\cal W}_{n_3}^2 + 2 {\cal W}^2 = {\cal V}
\ee
which again, coincides with the relation (the "superpotential equation") of Ref. \cite{gaugedBPS-bS}. In the present $N=2$ SUSY context, this relation may be easily understood from the fact that both ${\cal W}$ and ${\cal V}$ are derived from the same Kaehler potential $K$.

\section{Bogomol'nyi solitons in a gauged O(3) sigma model from $N=2$ SUSY }

As emphasized already, our method for the calculation of BPS equations for $N=2$ SUSY extended theories is completely general for chiral $N=2$ superfields with or without gauge interaction, therefore we may use it to study further models. Concretely, we want to employ it 
 to obtain the Bogomol'nyi equations of the gauged nonlinear sigma model originally analyzed in \cite{schr1}. We remark that the $N=2$ SUSY extension of this model in the O(3) formulation has already been discussed in \cite{trip}. The gauged non-linear sigma term results from the generalized Kaehler term
\be
{L}_2^g=\int d^2\theta d^2\bar{\theta} \; \mbox{ln}(1+\Phi^\dagger e^V{\Phi})
\ee
where now the target space metric (=the Kaehler metric) is the one of the CP(1) model and, therefore, the corresponding Kaehler potential is fixed. 
The resulting lagrangian is like in Eq. (\ref{L-2-g}), but for fixed $g_{u\bar u}= (1+u\bar u)^{-2}$.
Further, we need the $N=2$ extension of the Maxwell lagrangian, Eqs. (\ref{L-M-N2-1}) and (\ref{L-M-N2-2}).
Focusing on the $D$-dependent terms for the moment we find (for a general Kaehler potential $K(u\bar u)$)
\be
({L}_2^g+{L}_{M})\vert_D=D u\bar{u} K'+\frac{1}{2}D^2
\ee
(remember $K' \equiv \partial_{|u|^2} K$) with the solution
\be
D=- u\bar{u} K'
\ee
and, therefore,  the potential term contribution to the lagrangian is
\be
\mathcal{V}=\frac{1}{2} (u\bar{u} K')^2.
\ee
For the specific Kaehler potential of the CP(1) model, $K=\mbox{ln}(1+u\bar{u})$, we get
\be
\mathcal{V}=\frac{1}{2}\left(\frac{u\bar{u}}{1+u\bar{u}} \right)^2 .
\ee
We emphasize that this potential stems exclusively from the auxiliary field $D$, and that its form is fixed by the target space geometry (by the Kaehler potential). Specifically, there is no superpotential contribution to this potential, and the only solution for the auxiliary fields $F$ for this lagrangian is the trivial solution  $F=\bar{F}=0$. Using these solutions for $F$ and $D$, and setting the scalar $\sigma$ from the Maxwell superfield equal to its trivial solution, $\sigma =0$, we get the lagrangian in the bosonic sector
\be
({L}_2^g+{L}_{M})\vert_{bos}=\frac{D^\mu u D_\mu \bar u}{(1+u\bar u)^2 } - \frac{1}{2}\left(\frac{u\bar{u}}{1+u\bar{u}} \right)^2 -
\frac{1}{4} F_{\mu\nu}F^{\mu\nu}
\ee
that is, precisely the Lagrangian of the gauged nonlinear sigma model. Further,
inserting the on-shell values for the $D$ and $F$ fields into the general $N=2$ BPS equations (\ref{B1}), (\ref{B2}), we find the 
 Bogomol'nyi  equations 
\bea
D_1 u&=&\pm i D_2 u \label{Sc1}\\
B\equiv F_{12}&=&\pm \frac{\vert u \vert^2}{(1+\vert u \vert^2)}\label{Sc2},
\eea
which coincide precisely with the ones of Ref. \cite{schr1}. 

We remark that in this case, in principle, we may add a superpotential term
\be
L_0 = \int d^2 \theta {\cal U}(\Phi) + \int d^2 \bar \theta {\cal U}^\dagger (\Phi^\dagger),
\ee 
which leads to the $F$-dependent contribution
\be
(1+u\bar u)^{-2} F\bar F + {\cal U}_u F + {\cal U}^\dagger_{\bar u} \bar F
\ee
and to the on-shell values 
\be
\bar F = -(1+u\bar u)^2 {\cal U}_u \; , \quad F = - (1+u\bar u)^2 {\cal U}^\dagger_{\bar u}
\ee
and, therefore, to the further contribution to the potential
\be 
\tilde {\cal V} = (1+u\bar u)^4 |{\cal U}_u|^2 .
\ee
The BPS equations in this case read
\bea
F_{12}&=&\pm \frac{\vert u \vert^2}{(1+\vert u \vert^2)}\\
(1+u\bar u)^4 \vert {\cal U}_u \vert^2&=&D_iu D_i \bar{u}\pm\sqrt{(D_iu D_i \bar{u})^2-(D_i u)^2(D_j \bar{u})^2}.
\eea
The second BPS equation may be rewritten like
$$
2 (1+u\bar u)^4 \vert {\cal U}_u \vert^2 = (D_i u \pm i \epsilon_{ij} D_j u ) (D_i \bar{u} \mp i
\epsilon_{ik} D_k u )
$$
or (after introducing the complex base space variable $z = (1/2)(x+iy)$), depending on the sign, as
$$
2 (1+u\bar u)^4 \vert {\cal U}_u \vert^2 =( D_{\bar{z}} u ) ( D_z \bar{u} )
$$
or as
$$
2 (1+u\bar u)^4 \vert {\cal U}_u \vert^2 = ( D_z u ) (D_{\bar{z}} \bar{u} )
$$
where $\partial_z=\partial_x-i\partial_y$ and $A_z=A_x+iA_y$.

It might be interesting to investigate whether in this class of field theories some models (i.e., some nontrivial choices of ${\cal U}$) can be found which support 
genuine solitons.

\section{Conclusions}

It was the purpose of the present work to investigate in detail possible supersymmetric extensions of baby Skyrme models. First of all, we found that the complete baby Skyrme model, consisting of three terms (potential, quadratic and quartic term), allows for an $N=1$ SUSY extension where, in addition, the potential derives from a superpotential via the field equation of the auxiliary field, as usual. This finding is related to the fact that for this $N=1$ extension, the SUSY extension of the quartic term does not depend on the auxiliary field, at least in the bosonic sector. As a consequence, this SUSY extension cannot be used for the so-called BPS baby Skyrme model (a submodel without the quadratic term), because then the equation for the auxiliary field automatically eliminates the potential. Still, there exists another $N=1$ SUSY extension which automatically eliminates the quadratic term and induces the potential from the Kaehler metric (and {\em not} from a superpotential), leading directly to the BPS baby Skyrme model in the bosonic sector. It turns out that this $N=1$ extension is, in fact, secretly $N=2$. We explicitly constructed this $N=2$ extension and demonstrated that, again, the equation for the auxiliary field eliminates the quadratic term and induces the potential from the Kaehler metric. In a next step, we derived the general BPS equations for any $N=2$ supersymmertic field theory of chiral superfields and used this construction to demonstrate that the BPS solitons of the BPS baby Skyrme model are one-half BPS states of the corresponding $N=2$ supersymmetric extension. Then we turned to the investigation of SUSY extensions of gauged baby Skyrme models, i.e., of baby Skyrmions coupled to an abelian gauge field. We found that the complete gauged baby Skyrme model, too, has an $N=1$ extension. Further, the gauged BPS baby Skyrme model (without the quadratic term, but coupled to a gauge field) again has an $N=2$ extension where the auxiliary field of the chiral multiplet eliminates the quadratic term, whereas both auxiliary fields (from the chiral and the gauge multiplets) induce the potential in terms of the Kaehler potential. We derived the completely general BPS equations for any $N=2$ chiral multiplet coupled to an $N=2$ gauge multiplet and used this result to re-derive the BPS equations of the gauged BPS baby Skyrme model \cite{gaugedBPS-bS} as one-half BPS equations of the $N=2$ extension. Finally, we applied our general $N=2$ BPS equations to the gauged nonlinear sigma model as a further, concrete application. 

With these results at hand, the issue of possible applications and generalizations arises naturally. First of all, our BPS equations hold completely generally for {\em any} $N=2$ supersymmetric field theory of (gauged or ungauged) chiral superfields, so it can obviously be used to find BPS equations for other models. Baby Skyrmions as such have found some applications in brane cosmology \cite{sawa1}, so their supersymmetric extensions may be of interest in this context. Another interesting issue is related to generalizations to higher dimensions. An $N=2$ supersymmetric theory in $d=2+1$ dimensions leads in a natural way to an $N=1$ theory in one dimension higher, i.e., in $d=3+1$ dimensions. For the Skyrme-Faddeev-Niemi (SFN) model (same field content and lagrangian as the baby Skyrme model, but in $d=3+1$), we conclude that we cannot find an $N=1$ extension with our methods, in agreement with the findings of \cite{nepo}, \cite{frey}. On the other hand, for the restricted or extreme SFN model consisting of the quartic term and a potential only, we conclude that an $N=1$ SUSY extension does exist. This model has been investigated recently \cite{res-SFN1}, \cite{res-SFN2} where it was shown that it supports knotted and linked solitons (Hopfions), like the full SFN model.  In the same line of reasoning, we conclude that the gauged nonlinear sigma model in $d=3+1$ dimensions has an $N=1$ SUSY extension. 

This naturally leads to the question of SUSY extensions of the Skyrme model in $d=3+1$ dimensions. Indeed, the Skyrme model, too, has a submodel which supports BPS solitons \cite{BPS-Sk1}, and the results of the present work make it plausible to conjecture that this submodel might allow for an $N=2$ extension, as well, but now in $d=3+1$. This then implies that there should exist certain generalizations (i.e., more general submodels of the Skyrme model) which, while not possessing $N=2$ extensions, still allow for $N=1$ extensions. It would be very interesting to find these Skyrme submodels amenable to supersymmetry, to determine their SUSY extensions, and to investigate whether these supersymmetrizable Skyrme models are of special relevance in other contexts. 

Another interesting class of problems is related to (and requires the determination of) the fermionic sectors of the SUSY extensions of the non-standard kinetic terms. Due to their complexity, these fermionic sectors have remained undetermined in almost all calculations up to now. Their knowledge, however, would allow to determine explicitly the supercharges (not only their evaluation on BPS solutions) and to calculate the resulting SUSY algebra with its possible central extensions. It is well-known that in the presence of topological solitons these central extensions have to be expected \cite{witten-olive}. In addition, the inclusion of the fermions would allow to study the presence of fermionic zero modes in the background of topological solitons and, therefore, to investigate the corresponding index theorems relating the topological charges to the number of zero modes.  
These issues are under current investigation.

To summarize, in the present work we have made some important steps towards a better understanding of SUSY extensions of field theories with non-standard kinetic terms and, specifically, of non-standard field theories which support topological solitons. We found - among other results - that also for these theories the existence of BPS solitons is related to the existence of higher SUSY extensions, such that the BPS solitons are realized as BPS states in the SUSY extended theories, which for this type of theories is a new result. 

\section*{Acknowledgement}
The authors acknowledge financial support from the Ministry of Education, Culture and Sports, Spain (grant FPA2008-01177), the Xunta de Galicia (grant INCITE09.296.035PR and Conselleria de Educacion), the Spanish Consolider-Ingenio 2010 Programme CPAN (CSD2007-00042), and FEDER. Further, AW was supported by polish NCN grant 2011/01/ B/ST2/00464.


\begin{thebibliography}{100}
\bibitem{skyrme} T.H.R. Skyrme, Proc. Roy. Soc. Lon. {\bf 260},
127 (1961); Nucl. Phys. {\bf 31}, 556 (1962); J. Math. Phys. {\bf
12}, 1735 (1971).
\bb{AdNaWi}
G. Adkins, C. Nappi, E. Witten, Nucl. Phys. B{\bf 228}, 552 (1983).
\bb{AdNa}
G. Adkins, C. Nappi,  Nucl. Phys. B{\bf 233}, 109 (1984).
\bb{BraCar}
E. Braaten, L. Carson,
Phys. Rev. D{\bf 38}, 3525 (1988).
\bb{man1}
O.V. Manko, N.S. Manton, S.W. Wood,
Phys. Rev. C{\bf 76}, 055203 (2007) 
[arXiv:0707.0868];
R.A. Battye, N.S. Manton, P.M. Sutcliffe, S.W. Wood,
Phys. Rev. C{\bf 80}, 034323 (2009) 
[arXiv:0905.0099].
\bb{BaSu1}
R.A. Battye, P.M. Sutcliffe,
Phys. Rev. Lett. {\bf 79}, 363 (1997)
[hep-th/9702089]; 
Phys. Rev. Lett. {\bf 86}, 3989 (2001) 
[hep-th/0012215];
Rev.  Math .Phys. {\bf 14}, 29 (2002) 
[hep-th/0103026].
\bb{BaSu2}
R.A. Battye, P.M. Sutcliffe,
Nucl. Phys. B{\bf 705}, 384 (2005)
[hep-ph/0410157];
Phys. Rev. C{\bf 73}, 055205 (2006)
[hep-th/0602220].
\bb{man-sut-book}
N. Manton, P. Sutcliffe, "Topological Solitons", Cambridge University Press, Cambridge, 2007.
\bibitem{thooft} G. t'Hooft, Nucl. Phys. {\bf B72} (1974) 461; E. Witten, 
Nucl. Phys. {\bf B160} (1979) 57.   
\bibitem{witten1} E. Witten, Nucl. Phys. B223 (1983) 433.
\bibitem{old} B.M.A.G. Piette, B.J. Schoers and W.J.
Zakrzewski, Z. Phys. C {\bf 65} (1995) 165; B.M.A.G. Piette, B.J. 
Schoers and W.J. Zakrzewski, Nucl. Phys. B {\bf 439} (1995) 205.
\bibitem{holom} R.A. Leese, M. Peyrard and W.J. Zakrzewski
Nonlinearity {\bf 3} (1990) 773; B.M.A.G. Piette and W.J.
Zakrzewski, Chaos, Solitons and Fractals {\bf 5} (1995) 2495; P.M. 
Sutcliffe, Nonlinearity (1991) {\bf 4} 1109.
\bibitem{new} T. Weidig, Nonlinearity {\bf 12} (1999) 1489.
\bibitem{other} P. Eslami, M. Sarbishaei and W.J. Zakrzewski,
Nonlinearity {\bf 13} (2000) 1867.
\bibitem{kudr}
A.E. Kudryavtsev, B. Piette, W.J. Zakrzewski, Eur. Phys. J. C{\bf 1}, 333 (1998).
\bibitem{karliner} M. Karliner, I. Hen, Nonlinearity {\bf 21} (2008)
399-408;  M. Karliner, I. Hen, arXiv:0901.1489.
\bb{comp-bS}
C. Adam, P. Klimas, J. Sanchez-Guillen, A. Wereszczynski, Phys. Rev. D{\bf 80}, 105013 (2009).
\bb{jay1}
J. Jaykka, M. Speight, P. Sutcliffe, Proc. Roy. Soc. Lond. A{\bf 468}, 1085
(2012).
\bb{jay2}
J. Jaykka, M. Speight,  Phys. Rev. D{\bf 82},125030 (2010).
\bb{foster1}
D. Foster, Nonlinearity {\bf 23}, 465 (2010).
\bibitem{qhe} S.L. Sondhi, A. Karlhede, S.A. Kivelson, E.H.
Rezayi, Phys. Rev. B {\bf 47} (1993) 16419; N.R. Walet, T. Weidig,
Europhys. Lett. {\bf 55} (2001) 633.
\bb{sawa1}
Y. Kodama, K. Kokubu, N. Sawado, Phys. Rev. D{\bf 79}, 065024 (2009);
Y. Brihaye, T. Delsate, N. Sawado, Y. Kodama, Phys. Rev. D{\bf 82}, 106002 (2010);
T. Delsate, N. Sawado,  Phys. Rev. D{\bf 85}, 065025 (2012).
\bb{GP}
T. Gisiger, M.B. Paranjape, Phys. Rev. D{\bf 55}, 7731 (1997).
\bibitem{ward} M. de Innocentis, R. S. Ward, Nonlinearity {\bf 14} (2001) 663.
\bb{restr-bS}
C. Adam, T. Romanczukiewicz,  J. Sanchez-Guillen, A. Wereszczynski, 
Phys. Rev. D{\bf 81}, 085007 (2010).
\bb{Sp1}
J.M. Speight, 
J. Phys. A{\bf 43}, 405201 (2010). 
\bb{BPS-Sk1}
C. Adam, J. Sanchez-Guillen, A. Wereszczynski,
Phys. Lett. B{\bf 691}, 105 (2010)
[arXiv:1001.4544].
\bb{BPS-Sk2}
C. Adam, J. Sanchez-Guillen, A. Wereszczynski,
Phys. Rev. D{\bf 82}, 085015 (2010)  
[arXiv:1007.1567].
\bb{BPS-Sk-fosco}
C. Adam, C.D. Fosco, J.M. Queiruga, J. Sanchez-Guillen, A. Wereszczynski,
J. Phys. A {\bf 46}, 135401 (2013) [arXiv:1210.7839].
\bb{BoMa}
E. Bonenfant, L. Marleau,
Phys. Rev. D{\bf 82}, 054023 (2010)
[arXiv:1007.1396]; 
E. Bonenfant, L. Harbour, L. Marleau,
arXiv:1205.1414.
\bibitem{divecchia-ferrara}
P. DiVecchia, S. Ferrara, Nucl. Phys. B{\bf 130}, 93 (1977).
\bb{witten-olive}
E. Witten, D. Olive, Phys. Lett. B{\bf 78}, 97 (1978). 
\bb{d'adda-horsley}
A. D'Adda, R. Horsley, P. DiVecchia,
Phys. Lett B{\bf 76}, 298 (1978).
\bb{d'adda-divecchia}
P. DiVecchia, S. Ferrara,
Phys. Lett B{\bf 73}, 162 (1978).
\bb{HlSp1}
Z. Hlousek, D. Spector, Nucl. Phys. B {\bf 370} (1992) 143.
\bibitem{LeeLeeW}
C. Lee, K. Lee, E.J. Weinberg, Phys. Lett. B{\bf 243}, 105 (1990).
\bb{edelstein-nunez}
J.D. Edelstein, C. Nu\~nez, F. Schaposnik, Phys. Lett. B{\bf 329}, 39 (1994).
\bb{nepo}
E.A. Bergshoeff, R.I. Nepomechie, H.J. Schnitzer,  
Nucl. Phys. B{\bf 249}, 93 (1985).
\bb{frey}
L. Freyhult, 
Nucl. Phys. B{\bf 681}, 65 (2004) 
[hep-th/0310261].
\bb{bazeia-susy} 
D. Bazeia, R. Menezes,  A.Yu. Petrov, 
Phys. Lett. B{\bf 683}, 335 (2010) 
[arXiv:0910.2827].
\bibitem{susy-bS}
C. Adam, M. Queiruga, J. Sanchez Guillen, A. Wereszczynski,
Phys. Rev. D{\bf 84}, 025008 (2011)
[arXiv:1105.1168].
\bibitem{susy-def}
C. Adam, M. Queiruga, J. Sanchez Guillen, A. Wereszczynski,
Phys. Rev. D{\bf 84}, 065032 (2011)
[arXiv:1107.4370].
\bibitem{susy-def-bps}
C. Adam, M. Queiruga, J. Sanchez Guillen, A. Wereszczynski,
Phys. Rev. D{\bf 86}, 105009 (2012)
[arXiv:1209.6060].
\bb{ovrut1}
J. Khoury, J. Lehners, B. Ovrut, Phys. Rev. D{\bf 83} (2011) 125031 , arXiv:1012.3748.
\bb{ovrut2}
J. Khoury, J. Lehners, B. Ovrut, Phys. Rev. D{\bf 84} (2011) 043521 , arXiv:1103.0003.
\bb{ovrut3}
M. Koehn, J-L. Lehners, B. Ovrut,
Phys. Rev. D{\bf 86} (2012) 085019, 
arXiv:1207.3798.
\bb{FarKe1}
F. Farakos, C. Germani, A. Kehagias, E.N. Saridakis,
JHEP 1205 (2012) 050 
[arXiv:1202.3780]
\bb{FarKe2}
F. Farakos, A. Kehagias, 
 JHEP 1211 (2012) 077 
[arXiv:1207.4767].
\bb{nitta1}
L-X. Liu, M. Nitta, Int. J. Mod. Phys. A{\bf 27} (2012) 1250097.
\bb{nitta2}
M. Eto, T. Fujimori, M. Nitta, K. Ohashi, N.
Sakai, Prog. Theor. Phys. {\bf 128}  (2012) 67.
\bb{nitta3}
M. Nitta, arXiv:1206.5551.
\bb{adel1}
M. Dias, A. Yu. Petrov, C. R. Senise Jr., A. J. da Silva,
        arXiv:1212.5220 
\bb{k-infl}
C. Armendariz-Picon, T. Damour,  V. Mukhanov, Phys. Lett. B{\bf458}, 209 (1999) [hep-th/9904075].
\bb{k-ess}C. Armendariz-Picon, V. Mukhanov, P.J. Steinhardt, Phys. Rev. Lett. {\bf85}, 4438 (2000) [astro-ph/0004134];
C. Armendariz-Picon, V. Mukhanov, P.J. Steinhardt, Phys. Rev. D{\bf63}, 103510 (2001) [astro-ph/0006373].
\bb{bab-muk-1}
E. Babichev, V. Mukhanov, A. Vikman,
JHEP {\bf 0802}, 101 (2008) 
[arXiv:0708.0561]
\bb{babichev1} E. Babichev, Phys. Rev. D{\bf 74}, 085004 (2006) [hep-th/0608071].
\bb{comp-brane} C. Adam, N. Grandi, J. Sanchez-Guillen, A. Wereszczynski, J. Phys. A{\bf41}, 212004 (2008) [arXiv:0711.3550];
C. Adam, N. Grandi, P. Klimas, J. Sanchez-Guillen, A. Wereszczynski, J. Phys. A{\bf41}, 375401 (2008) [arXiv:0805.3278].
\bb{olech}M. Olechowski, Phys. Rev. D{\bf78}, 084036 (2008) [arXiv:0801.1605].
\bb{bazeia3}D. Bazeia, L. Losano, R. Menezes, Phys. Lett. B{\bf 668}, 246 (2008), [arXiv:0807.0213]; D. Bazeia, A. R. Gomes, L. Losano, R. Menezes, Phys. Lett. B{\bf 671}, 402 (2009) [arXiv:0808.1815].
\bb{liu}
Y.-X. Liu, Y. Zhong, K. Yang, Europhys. Lett. {\bf 90}, 51001 (2010).
\bb{trodden}
M. Andrews, M. Lewandowski, M. Trodden, D. Wesley, Phys. Rev. D{\bf 82}, 105006 (2010).
\bb{Dzhu1}
V. Dzhunushaliev, V. Folomeev, M. Minamitsuji, 
Rept. Prog. Phys. {\bf 73}, 066901 (2010), 
[arXiv:0904.1775].
\bb{GPS}
J. Gladikowski, B.M.A.G. Piette, B.J. Schroers,
Phys. Rev. D{\bf 53} 844, 1996
[hep-th/9506099].
\bb{gaugedBPS-bS}
C. Adam, C. Naya, J. Sanchez-Guillen, A. Wereszczynski,
Phys. Rev. D{\bf 86} (2012) 045010 
[arXiv:1205.1532]. 
\bb{schr1}
B.J. Schroers,
Phys. Lett. B{\bf 356}  291, 1995
[hep-th/9506004].
\bb{trip}
P.K. Tripathy,
Phys. Rev. D{\bf 59} (1999) 085004 
[hep-th/9811186]. 
\bb{res-SFN1}
C. Adam, J. Sanchez-Guillen, T. Romanczukiewicz, A. Wereszczynski, 
J. Phys. A{\bf 43} (2010) 345402 
[arXiv:0911.3673].
\bb{res-SFN2}
D. Foster,
Phys. Rev. D{\bf 83} (2011) 085026 
[arXiv:1012.2595]. 

\end{thebibliography}
\end{document}